\documentclass[reprint, superscriptaddress, twocolumn, amsmath, amssymb, aps, prb]{revtex4-2}
\usepackage{graphicx}
\usepackage{dcolumn}
\usepackage{bm}
\usepackage{placeins}
\usepackage{appendix}
\usepackage{braket}

\usepackage{upgreek}

\usepackage{verbatim}
\usepackage[usenames,dvipsnames]{xcolor}
 \usepackage{amsmath}
 \usepackage{amsfonts}
 \usepackage{amssymb}
 \usepackage[colorlinks=true,citecolor=blue,linkcolor=red]{hyperref}

 \usepackage{bbold}     
 \usepackage[makeroom]{cancel}  
 \usepackage{multirow}    
 \usepackage[normalem]{ulem}        
 \usepackage{array}
 \usepackage{pdfpages}
\makeatletter
 \AtBeginDocument{\let\LS@rot\@undefined}
 \makeatother

\begin{document}


\title{Gatemon qubit based on a thin InAs-Al hybrid nanowire}

\author{Jierong Huo}
 \email{equal contribution}
\affiliation{State Key Laboratory of Low Dimensional Quantum Physics, Department of Physics, Tsinghua University, Beijing 100084, China}

\author{Zezhou Xia}
\email{equal contribution}
\affiliation{State Key Laboratory of Low Dimensional Quantum Physics, Department of Physics, Tsinghua University, Beijing 100084, China}

\author{Zonglin Li}
 \email{equal contribution}
\affiliation{State Key Laboratory of Low Dimensional Quantum Physics, Department of Physics, Tsinghua University, Beijing 100084, China}

\author{Shan Zhang}
 \email{equal contribution}
\affiliation{State Key Laboratory of Low Dimensional Quantum Physics, Department of Physics, Tsinghua University, Beijing 100084, China}

\author{Yuqing Wang}
\affiliation{Beijing Academy of Quantum Information Sciences, Beijing 100193, China}

\author{Dong Pan}
\affiliation{State Key Laboratory of Superlattices and Microstructures, Institute of Semiconductors, Chinese Academy of Sciences, P. O. Box 912, Beijing 100083, China}

\author{Qichun Liu}
\affiliation{Beijing Academy of Quantum Information Sciences, Beijing 100193, China}

\author{Yulong Liu}
\affiliation{Beijing Academy of Quantum Information Sciences, Beijing 100193, China}

\author{Zhichuan Wang}
\affiliation{Beijing National Laboratory for Condensed Matter Physics, Institute of Physics, Chinese Academy of Sciences, Beijing 100190, China}

\author{Yichun Gao}
\affiliation{State Key Laboratory of Low Dimensional Quantum Physics, Department of Physics, Tsinghua University, Beijing 100084, China}

\author{Jianhua Zhao}
\affiliation{State Key Laboratory of Superlattices and Microstructures, Institute of Semiconductors, Chinese Academy of Sciences, P. O. Box 912, Beijing 100083, China}

\author{Tiefu Li}
\affiliation{School of Integrated Circuits and Frontier Science Center for Quantum Information, Tsinghua University, Beijing 100084, China}
\affiliation{Beijing Academy of Quantum Information Sciences, Beijing 100193, China}

\author{Jianghua Ying}
 \email{yingjianghua@tgqs.net}
\affiliation{Yangtze Delta Region Industrial Innovation Center of Quantum and Information, Suzhou 215133, China}

\author{Runan Shang}
\affiliation{Beijing Academy of Quantum Information Sciences, Beijing 100193, China}

\author{Hao Zhang}
\email{hzquantum@mail.tsinghua.edu.cn}
\affiliation{State Key Laboratory of Low Dimensional Quantum Physics, Department of Physics, Tsinghua University, Beijing 100084, China}
\affiliation{Beijing Academy of Quantum Information Sciences, Beijing 100193, China}
\affiliation{Frontier Science Center for Quantum Information, Beijing 100084, China}


\begin{abstract}

We study a gate-tunable superconducting qubit (gatemon) based on a thin InAs-Al hybrid nanowire. Using a gate voltage to control its Josephson energy, the gatemon can reach the strong coupling regime to a microwave cavity. In the dispersive regime, we extract the energy relaxation time $T_1\sim$0.56  $\upmu$s and the dephasing time $T_2^* \sim$0.38 $\mathrm{\upmu}$s. Since thin InAs-Al nanowires can have fewer or single sub-band occupation and recent transport experiment shows the existence of nearly quantized zero-bias conductance peaks, our result holds relevancy for detecting Majorana zero modes in thin InAs-Al nanowires using circuit quantum electrodynamics.

\end{abstract}

\maketitle

Topological quantum computation \cite{Kitaev_toric, TQC_RMP} aims to solve the decoherence problem at the device level by encoding information into Majorana zero modes \cite{ReadGreen, Kitaev}. A promising material candidate is the semiconductor-superconductor hybrid nanowires \cite{Lutchyn2010, Oreg2010}. Tremendous efforts have been put into searching for possible Majorana signatures in InAs and InSb nanowires \cite{Mourik, Deng2016, Gul2018, Zhang2021, Song2022, WangZhaoyu, Prada2020,NextSteps}. Meanwhile, proposals on topological qubits have been theoretically explored with great enthusiasm \cite{Alicea_NP, Hyart_PRB, Dong_PRX_2016, Alicea_PRX, Plugge_2017, Fu_braiding, MS_scalable}. A major technique in those proposals is the circuit quantum electrodynamics (cQED), similar to that in the superconducting transmon qubit \cite{Transmon, Transmon_Nature}. Moreover, cQED could also be used to probe Majorana signatures if incorporating the nanowire into a transmon-like device \cite{Majorana_transmon, 2015_PRB_MZM_transmon, 2018_PRB_MZM_transmon, 2020_PRR_MZM_transmon, 2022_PRL_MZM_transmon}. Motivated by this, transmon qubits based on InAs-Al nanowires have been realized and studied in recent years \cite{2015_PRL_gatemon, DiCarlo_2015, Gatemon_2016_PRL, DiCarlo_2018, 2020_PRL_Charlie_transmon, 2020_PRL_Leo_transmon, 2020_PRL_pi_qubit, 2020_PRL_fullshell, 2022_PRXQuantum}. The InAs wire diameters in those gate-tunable transmons (gatemons) are typically large, $\sim$75-160 nm. Though the junction region can be easily gate-tuned, the proximitized InAs region is heavily screened by the covered Al film and is still in the multi-subband regime. Thick wire and multi-band bring challenges into the Majorana detection \cite{Caroff2009, Shtrikman2009, Pan2014, Brouwer2012ZBP, Loss2013ZBP, GoodBadUgly}. To overcome this issue, thin InAs-Al nanowires have been explored and nearly quantized zero bias conductance peaks have been reported \cite{PanCPL, Song2022, WangZhaoyu}. Here, we report the realization of gatemon qubit based on these thin wires. The InAs diameter is $\sim$35 nm, significantly smaller than those in previous gatemons. Our result paves the way for future Majorana cQED experiments. 

\begin{figure*}[htb]
\includegraphics[width=0.9\textwidth]{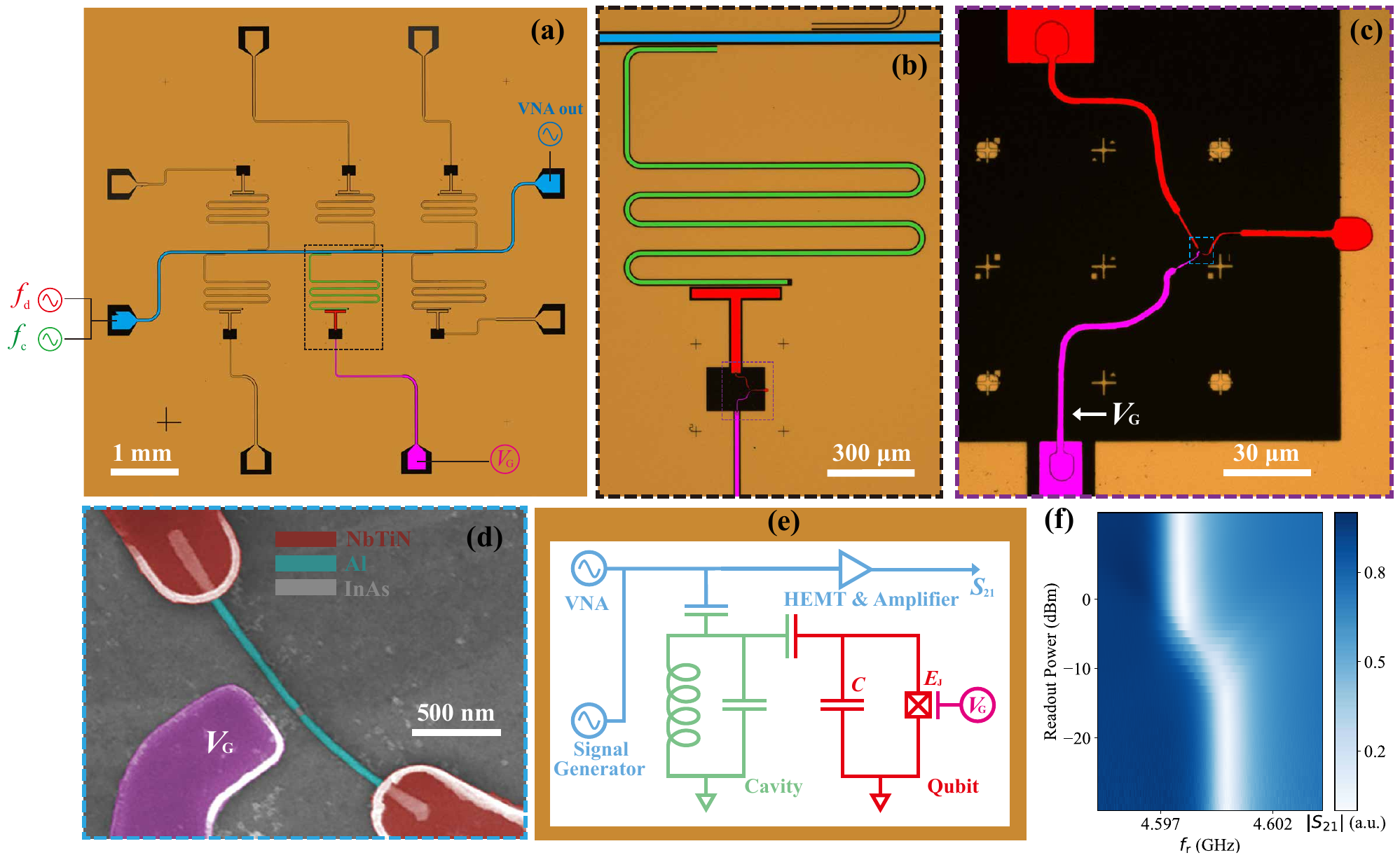}
\centering
\caption{(a) Optical image of a gatemon chip and measurement set-up. The grounding plane is in orange. The common feed line is in blue (false colored). The $\lambda/4$ cavity (for device A) is in green. The T-shape qubit capacitor (for device A) is in red. The gate line is in pink. All these elements above are NbTiN (thickness 100 nm) and were fabricated in one lithography step using reactive ion etching. (b) An enlargement of (a) (the dashed box). (c) SEM (false colored) of the InAs-Al region of the gatemon (the dashed box in (b)). The top contact connects to the T-shape capacitor. The bottom contact is grounded. A side gate (pink) connects to the gate line. The gate and contacts were fabricated in another lithography step by sputtering Ti/NbTiN (1/100 nm). (d) SEM of the Josephson junction region of the qubit (the blue box in (c)). (e) Schematic of the measurement circuit. (f) Feed line transmission as a function of the readout power. $V_{\text{G}}$ = -4.554 V. The cavity shift indicates the coupling of the cavity to a nonlinear circuit (the qubit). }
\label{fig1}
\end{figure*}

\textbf{Qubit device and measurement circuit.} Figure 1(a) shows the optical image (false colored) of the device chip. One gatemon qubit (device A, the dashed box) was measured with results shown in Figs. 2-4. The other five qubits on this chip were not working. We have characterized three working qubits, see the supplementary material (SM) for the other two. For the device fabrication, a 100-nm-thick NbTiN superconducting film (the orange region) was first sputtered onto a sapphire substrate. Reactive ion etching was then performed to etch away part of the film (the dark regions).  This lithography step defines the co-planar wave-guide feed line (blue), the resonator/cavity (green) and the shunt capacitor of the gatemon (red). Figure 1(b) is a zoomed-in image of the qubit. The feed line capacitively couples to a $\lambda/4$ cavity for the qubit readout. The cavity internal quality factor, $Q_{i}$, is $\sim$14000 and the bare resonance frequency $f_{\text{C}}$ is $\sim$4.6 GHz. The cavity further couples to the T-shape capacitor whose capacitance is estimated to be $\sim$100 fF. This capacitor connects to the InAs-Al nanowire Josephson junction (Figs. 1(c) and 1(d)) and together, they form the gatemon qubit. The large capacitance determines the charging energy to $E_\text{C} \sim$$e^2/2C \sim$190 MHz. The other side of the Josephson junction connects to the ground. A side gate (pink) tunes the junction transparency and therefore controls the Josephson energy, $E_{\text{J}}$. The junction was defined by removing (etching) a small Al segment on the InAs wire (diameter $\sim$35 nm). Transports on these wires show a gate-tunable supercurrent ($I_{\text{c}}$) on the order of 100 nA \cite{Zhichuan}. The corresponding $E_\text{J}=\hbar I_{\text{c}}/2e \sim$50 GHz is much larger than $E_\text{C}$. This ensures that the qubit can be operated in the transmon regime. For details of the device fabrication, circuit set-up and cavity calibration, see SM (the method session, Figs. S1 and S2). 

\begin{figure*}[htb]
\includegraphics[width=0.8\textwidth]{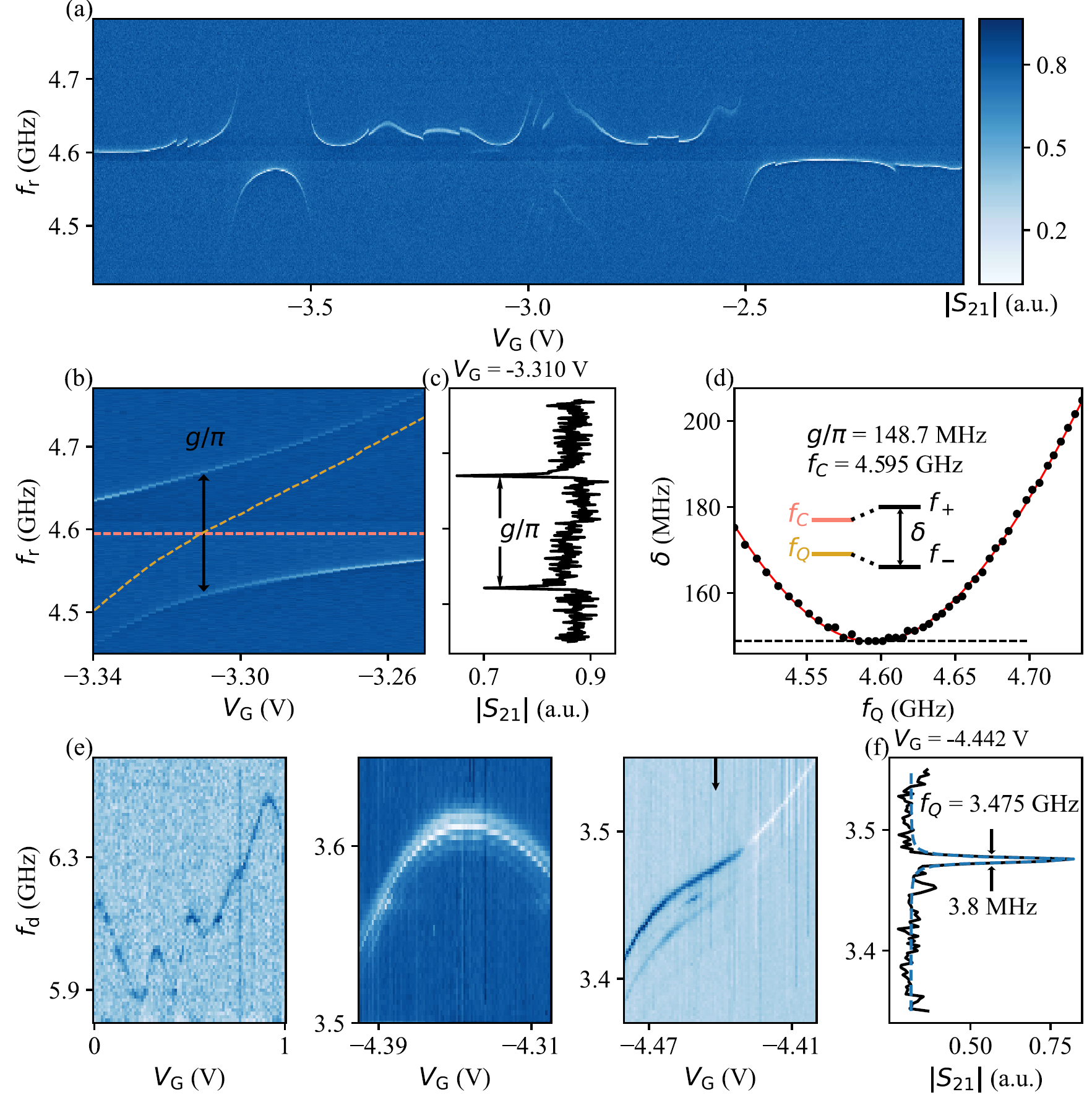}
\centering
\caption{(a) Cavity transmission in the single-tone measurement as a function of the cavity drive frequency and $V_{\text{G}}$. The anti-crossings are the vacuum Rabi splittings. (b) Fine scan of an anti-crossing. The gate voltage has a shift compared to the same feature in (a) due to hysteresis or charge jumps. The two dashed lines are the bare cavity frequency (pink, $f_{\text{C}}$) and the extracted qubit frequency (yellow, $f_{\text{Q}}$). (c) A line cut from (b) at the crossing point of the two dashed lines. The peak spacing $\delta = f_+ - f_-$ is indicated by the black arrow in (b). (d) $\delta$ as a function of the qubit frequency. The red line is the theoretical fit. Inset, energy schematic of the cavity-qubit hybridization and the parameters. (e) Gatemon qubit spectroscopy (two-tone) as a function of $V_{\text{G}}$ at three different ranges. (f) A line cut of (e) (at the black arrow) with a Lorentzian fit (blue dashed line).}
\label{fig2}
\end{figure*}

Figure 1(e) draws the equivalent circuit diagram. For the cavity and qubit readout, a microwave tone of frequency $f_{\text{r}}$ (near the cavity resonant frequency $f_{\text{C}}$) is applied to the feed line (see Fig. 1(a)). The transmission of this microwave tone, $S_{21}$, is measured by a vector network analyzer (VNA). Figure 1(f) is such a ``single-tone'' measurement  by sweeping $f_{\text{r}}$ and its power while monitoring the transmission amplitude $|S_{21}|$. At high power, the qubit is ``overwhelmed'' and the dip in $|S_{21}|$ corresponds to the bare resonant frequency of the cavity $f_{\text{C}}$ \cite{PRL_Reed}. At low power, the resonant frequency is qubit-state dependent due to the cavity-qubit dispersive interaction. The repulsion of the qubit and cavity causes the shift of the resonance frequency as shown in Fig. 1(f). To excite and control the qubit, a second microwave tone of frequency $f_{\text{d}}$ can be applied in the standard ``two-tone'' spectroscopy.

\textbf{Vacuum Rabi splitting and qubit spectroscopy.} In Fig. 2(a), we keep the readout power low and scan the gate voltage ($V_{\text{G}}$). The resonant frequency of the cavity is gate tunable, indicating the presence of the gatemon. For better visibility, a signal background, contributed by standing waves in the circuit, was subtracted from $|S_{21}|$ (see Fig. S3 in SM for details). The jumps in the spectrum are due to charge instabilities in the InAs-Al devices which are also commonly observed in the transport characterizations. The qubit frequency ($f_{\text{Q}}$) is given by the energy difference between the ground state and the first excited state, $hf_{\text{Q}}=E_{01}\sim$$\sqrt{8E_{\text{C}}E_{\text{J}}}-E_{\text{C}}$. $V_{\text{G}}$ tunes $E_{\text{J}}$ therefore controls $f_{\text{Q}}$. When $f_{\text{Q}}$ is tuned close to the cavity frequency ($f_{\text{C}}$), the strong qubit-cavity hybridization leads to the anti-crossings. This anti-crossing is observed in the single-tone spectrum shown in Fig. 2(a). See Fig. 2(b) for a fine scan (an enlargement) of an anti-crossing. Figure 2(c) is the line cut at $V_{\text{G}}$ = -3.310 V where the peak spacing is the smallest. In this strong coupling regime,  the frequencies of the two peaks are $f_{\pm}=\left[f_{\text{Q}}+f_{\text{C}}\pm \sqrt{(f_{\text{Q}}-f_{\text{C}})^2+4(g/2\pi)^2 }\right]/2$. The peak spacing, $\delta=f_{+}-f_{-}= \sqrt{(f_{\text{Q}}-f_{\text{C}})^2+4(g/2\pi)^2 }$, is a function of $f_{\text{Q}}$. The qubit frequency can be obtained by $f_{\text{Q}}=f_++f_--f_{\text{C}}$. Figure 2(d) plots the extracted $\delta$ and $f_{\text{Q}}$. The red line is the theory fit based on the formula above.  We extract the qubit-cavity coupling strength $g/2\pi \sim$74 MHz.

Next we tune $V_{\text{G}}$ to bias $f_{\text{Q}}$ away from $f_{\text{C}}$ and reach the dispersive regime. The large detuning, $|\Delta/2\pi| = |f_{\text{Q}}-f_{\text{C}} | \gg g$, could effectively suppress the energy relaxation due to the Purcell effect \cite{Purcell}. Figure 2(e) shows the two-tone spectroscopy, $|S_{21}|$ as a function of the qubit drive $f_{\text{d}}$ (the second tone) at three different $V_{\text{G}}$ ranges. The readout frequency $f_{\text{r}}$ (the first tone) was fixed near the cavity resonance ($f_{\text{C}}$). When $f_{\text{d}}$ is scanned on resonance with $f_{\text{Q}}$, the qubit can be excited and the resonator frequency is shifted. A signal in the readout tone can be observed due to the cavity shift. The spectroscopy in Fig. 2(e) reveals the gate tunable nature of $f_{\text{Q}}$. The non-monotonic fluctuations are associated with the non-ballistic property of the InAs-Al junction, indicating the presence of disorder.

\begin{figure*}[htb]
\includegraphics[width=0.8\textwidth]{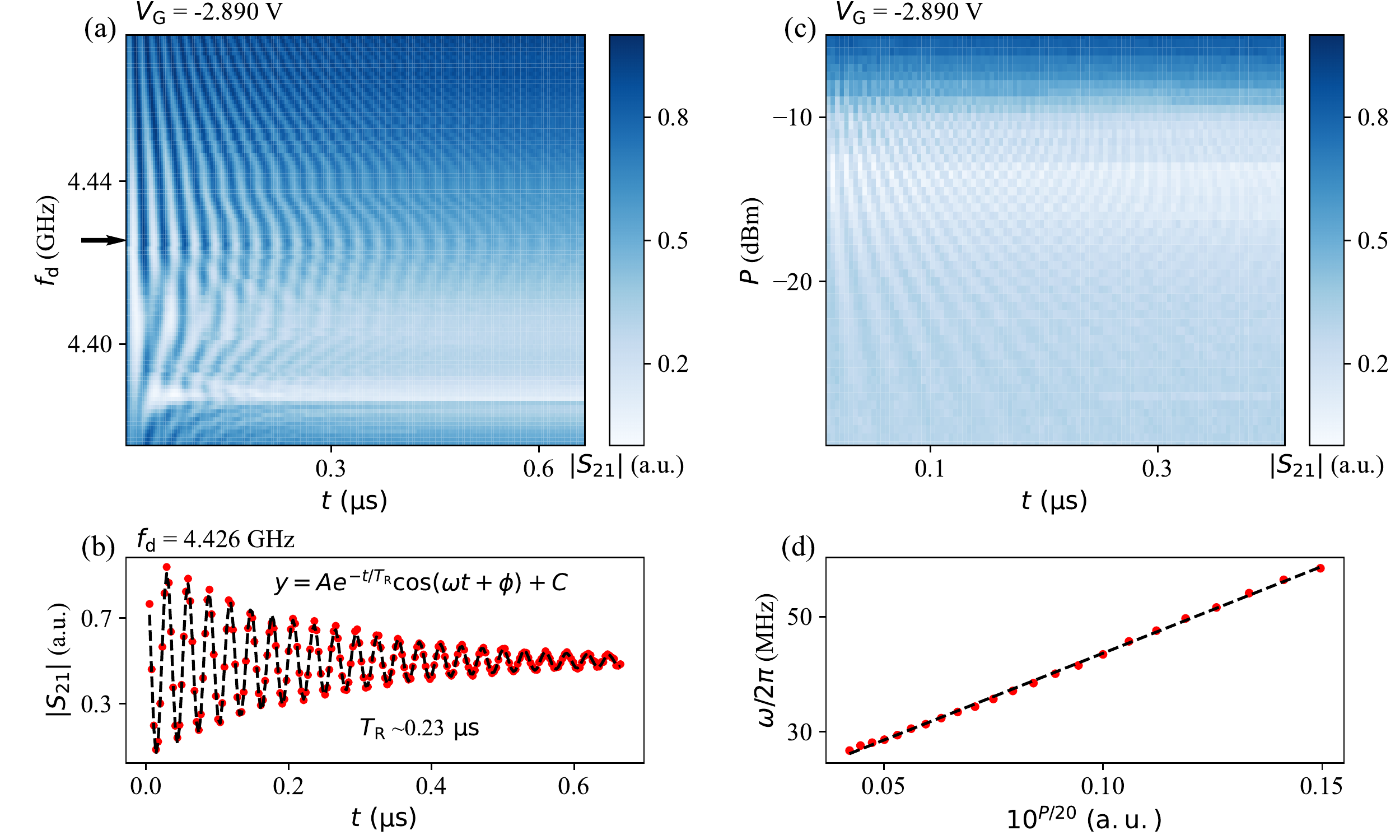}
\centering
\caption{(a) Rabi oscillations as a function of the qubit drive frequency and the drive pulse duration. (b) A line cut at the qubit resonant frequency $f_{\text{Q}}$ =  4.426 GHz (see the arrow in (a)). The dashed line is a theoretical fit. (c) Rabi oscillations as a function of the drive power at a fixed drive frequency $f_{\text{d}}=4.446$ GHz. (d) The Rabi oscillation frequency ($\omega/2\pi$) extracted from (c) versus the driving amplitude (over the range where the oscillations are visible). $10^{P/20}$ has a linear relation with the driving amplitude. The dashed line is a linear fit. $V_{\text{G}}=-2.890$ V for all panels.}
\label{fig3}
\end{figure*}

\begin{figure*}[htb]
\includegraphics[width=0.95\textwidth]{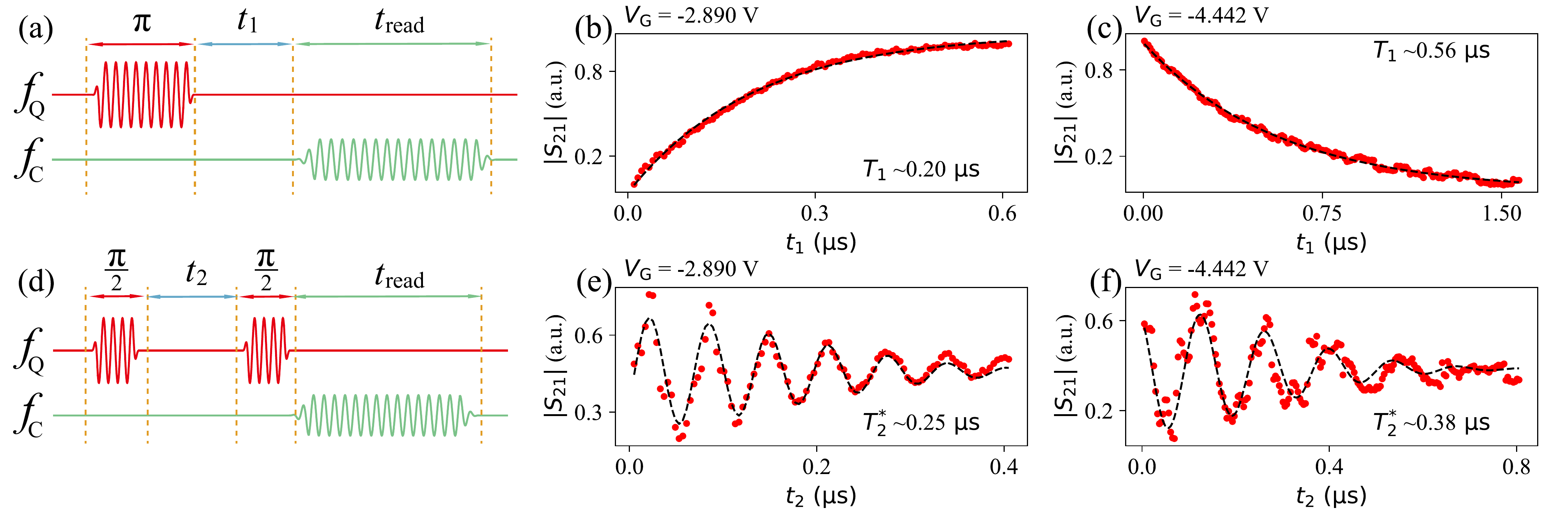}
\centering
\caption{Energy relaxation time $T_1$ and dephasing time $T_2^*$. (a) Pulse sequence schematic for the $T_1$ measurement. (b)-(c) $T_1$ measurements at $V_{\text{G}}$ of -2.890 V and -4.442 V, respectively. The black dashed lines are the exponential decay fits. The increasing vs decreasing trend between (b) and (c) are due to different selected working points, resulting in reversed readout signal strength. (d) Pulse sequence schematic for the $T_2^*$ measurement. (e)-(f) Ramsey oscillations at different $V_{\text{G}}$ values. The dashed lines are fits with an exponential decaying envelope. }
\label{fig4}
\end{figure*}

The multiple peaks in Fig. 2(e) are likely caused by the photon-number-dependent frequency shift of the qubit \cite{Photon_number}. This is obvious when the cavity readout tone was continuously applied (instead of pulsed) on the feed line, see Fig. S4 in SM for detailed analysis. The extracted peak spacing ($\sim$12 MHz) roughly matches our estimation of $2|\chi|/2\pi=g^2/|\Delta|\pi \sim$10 MHz. Figure 2(f) shows a line cut of the qubit excitation peak from Fig. 2(e). From the Lorentzian fit we extract the full width at half maximum (FWHM $\sim$3.8 MHz). This width corresponds to a coherence time $\sim$0.26 $\upmu$s, consistent with the time-domain measurement in Figs. 3 and 4.

\textbf{Rabi oscillations.} We now manipulate the gatemon qubit in time domain. A qubit drive pulse ($f_{\text{d}}$) was first applied for a duration time of $t$ and then followed by a readout pulse of the cavity. Figure 3(a) shows the typical Rabi oscillations as a function of $f_{\text{d}}$ and $t$. The Rabi oscillation frequency $\omega \propto \sqrt{(f_{\text{d}}-f_{\text{Q}})^2+{\text{const.}}}$. The term, const., is proportional to the square of the driving amplitude. The oscillation pattern shows a Chevron feature. We note that the pattern gets disrupted near 4.40 GHz, probably due to the presence of another cavity on the chip whose resonant frequency is around 4.40 GHz. From Fig. 3(a), we can estimate the qubit frequency $f_{\text{Q}} \sim$4.426 GHz. Note that this is not in the dispersive regime yet since the detuning is not large enough. Figure 3(b) shows the line cut near $f_{\text{Q}}$. Fitting the decaying oscillation using an empirical formula $y=Ae^{-t/T_{\text{R}}}\text{cos}(\omega t+\phi)+C$ (the dashed line), we extract the Rabi coherence time $T_{\text{R}}\sim$0.23 $\upmu$s.

In Fig. 3(c), the qubit drive power is varied. Higher power drives the qubit faster, resulting in a shorter oscillation period. We extract the oscillation frequency (the inverse of the period) and plot it as a function of its driving amplitude (converted from the power) in Fig. 3(d). The dashed line is a linear fit, confirming its Rabi oscillation nature.

\textbf{Gatemon quantum coherence.} To extract the gatemon energy relaxation time $T_1$, a $\pi$ pulse was first applied to excite the qubit to the $\ket{1}$ state.  The readout was performed after the waiting time $t_1$ (Fig. 4(a)). Figures 4(b) and 4(c) show the exponential fit at two different gate voltages. A relaxation time $T_1 \sim$0.56 $\upmu$s can be extracted. The dephasing time $T_2^*$ was determined by the Ramsey experiment: inserting a waiting time $t_2$ between two slightly detuned $\pi$/2 pulses before the readout (Fig. 4(d)). Figures 4(e) and 4(f) show two Ramsey oscillations. $T_2^*$ can reach $\sim$0.38 $\upmu$s. The fitting assumes an exponential decaying envelope: $A\text{cos}(\omega t+\phi)\text{exp}(-(t_2/T_2^*)^2)+C$ \cite{guide}. Note that the gate voltage of Figs. 4(c) and 4(f) corresponds to the dispersive regime ($f_{\text{Q}}$ here is $\sim$3.446 GHz), therefore a longer coherence time is expected. As a comparison, the coherence times for Figs. 4(b) and 4(e) are shorter due to the Purcell effect. $T_2^*<2T_1$ indicates that the coherence of our qubit is not entirely limited by energy relaxation. For the Rabi oscillations at $V_{\text{G}}$ = -4.442 V (same with Figs. 4(c) and 4(f)), see Fig. S5. In Fig. S6, we show the measurement of two more gatemon qubits.

In summary, we have studied the gatemon qubit based on a thin InAs-Al hybrid nanowire. The gatemon can reach strong coupling to a cavity. Coherent Rabi oscillations can be observed. The qubit relaxation time $T_1$ and dephasing time $T_2^*$ can reach 0.56 $\upmu$s and 0.38 $\upmu$s, respectively. Future work on these thin-wire-based gatemons could aim for possible Majorana signatures in a finite magnetic field.

\textbf{Acknowledgment} We thank Chunqing Deng and Luyan Sun for valuable discussions. We also thank the Teaching Center for Experimental Physics of Tsinghua University for using their equipment. This work is supported by Tsinghua University Initiative Scientific Research Program, Alibaba Innovative Research Program, National Natural Science Foundation of China (Grant Nos. 12204047, 92065106, 61974138). D.P. also acknowledges the support from Youth Innovation Promotion Association, Chinese Academy of Sciences (Nos. 2017156 and Y2021043). Raw data and processing codes within this paper are available at https://doi.org/10.5281/zenodo.7620737. 

\bibliography{mybibfile}

\begin{thebibliography}{49}%
\makeatletter
\providecommand \@ifxundefined [1]{%
 \@ifx{#1\undefined}
}%
\providecommand \@ifnum [1]{%
 \ifnum #1\expandafter \@firstoftwo
 \else \expandafter \@secondoftwo
 \fi
}%
\providecommand \@ifx [1]{%
 \ifx #1\expandafter \@firstoftwo
 \else \expandafter \@secondoftwo
 \fi
}%
\providecommand \natexlab [1]{#1}%
\providecommand \enquote  [1]{``#1''}%
\providecommand \bibnamefont  [1]{#1}%
\providecommand \bibfnamefont [1]{#1}%
\providecommand \citenamefont [1]{#1}%
\providecommand \href@noop [0]{\@secondoftwo}%
\providecommand \href [0]{\begingroup \@sanitize@url \@href}%
\providecommand \@href[1]{\@@startlink{#1}\@@href}%
\providecommand \@@href[1]{\endgroup#1\@@endlink}%
\providecommand \@sanitize@url [0]{\catcode `\\12\catcode `\$12\catcode
  `\&12\catcode `\#12\catcode `\^12\catcode `\_12\catcode `\%12\relax}%
\providecommand \@@startlink[1]{}%
\providecommand \@@endlink[0]{}%
\providecommand \url  [0]{\begingroup\@sanitize@url \@url }%
\providecommand \@url [1]{\endgroup\@href {#1}{\urlprefix }}%
\providecommand \urlprefix  [0]{URL }%
\providecommand \Eprint [0]{\href }%
\providecommand \doibase [0]{https://doi.org/}%
\providecommand \selectlanguage [0]{\@gobble}%
\providecommand \bibinfo  [0]{\@secondoftwo}%
\providecommand \bibfield  [0]{\@secondoftwo}%
\providecommand \translation [1]{[#1]}%
\providecommand \BibitemOpen [0]{}%
\providecommand \bibitemStop [0]{}%
\providecommand \bibitemNoStop [0]{.\EOS\space}%
\providecommand \EOS [0]{\spacefactor3000\relax}%
\providecommand \BibitemShut  [1]{\csname bibitem#1\endcsname}%
\let\auto@bib@innerbib\@empty
\bibitem [{\citenamefont {Kitaev}(2003)}]{Kitaev_toric}%
  \BibitemOpen
  \bibfield  {author} {\bibinfo {author} {\bibfnamefont {A.}~\bibnamefont
  {Kitaev}},\ }\bibfield  {title} {\bibinfo {title} {Fault-tolerant quantum
  computation by anyons},\ }\href
  {https://doi.org/https://doi.org/10.1016/S0003-4916(02)00018-0} {\bibfield
  {journal} {\bibinfo  {journal} {Annals of Physics}\ }\textbf {\bibinfo
  {volume} {303}},\ \bibinfo {pages} {2} (\bibinfo {year} {2003})}\BibitemShut
  {NoStop}%
\bibitem [{\citenamefont {Nayak}\ \emph {et~al.}(2008)\citenamefont {Nayak},
  \citenamefont {Simon}, \citenamefont {Stern}, \citenamefont {Freedman},\ and\
  \citenamefont {Das~Sarma}}]{TQC_RMP}%
  \BibitemOpen
  \bibfield  {author} {\bibinfo {author} {\bibfnamefont {C.}~\bibnamefont
  {Nayak}}, \bibinfo {author} {\bibfnamefont {S.~H.}\ \bibnamefont {Simon}},
  \bibinfo {author} {\bibfnamefont {A.}~\bibnamefont {Stern}}, \bibinfo
  {author} {\bibfnamefont {M.}~\bibnamefont {Freedman}},\ and\ \bibinfo
  {author} {\bibfnamefont {S.}~\bibnamefont {Das~Sarma}},\ }\bibfield  {title}
  {\bibinfo {title} {Non-abelian anyons and topological quantum computation},\
  }\href {https://doi.org/10.1103/RevModPhys.80.1083} {\bibfield  {journal}
  {\bibinfo  {journal} {Rev. Mod. Phys.}\ }\textbf {\bibinfo {volume} {80}},\
  \bibinfo {pages} {1083} (\bibinfo {year} {2008})}\BibitemShut {NoStop}%
\bibitem [{\citenamefont {Read}\ and\ \citenamefont {Green}(2000)}]{ReadGreen}%
  \BibitemOpen
  \bibfield  {author} {\bibinfo {author} {\bibfnamefont {N.}~\bibnamefont
  {Read}}\ and\ \bibinfo {author} {\bibfnamefont {D.}~\bibnamefont {Green}},\
  }\bibfield  {title} {\bibinfo {title} {Paired states of fermions in two
  dimensions with breaking of parity and time-reversal symmetries and the
  fractional quantum hall effect},\ }\href
  {https://doi.org/10.1103/PhysRevB.61.10267} {\bibfield  {journal} {\bibinfo
  {journal} {Phys. Rev. B}\ }\textbf {\bibinfo {volume} {61}},\ \bibinfo
  {pages} {10267} (\bibinfo {year} {2000})}\BibitemShut {NoStop}%
\bibitem [{\citenamefont {Kitaev}(2001)}]{Kitaev}%
  \BibitemOpen
  \bibfield  {author} {\bibinfo {author} {\bibfnamefont {A.~Y.}\ \bibnamefont
  {Kitaev}},\ }\bibfield  {title} {\bibinfo {title} {Unpaired majorana fermions
  in quantum wires},\ }\href {https://doi.org/10.1070/1063-7869/44/10s/s29}
  {\bibfield  {journal} {\bibinfo  {journal} {Physics-Uspekhi}\ }\textbf
  {\bibinfo {volume} {44}},\ \bibinfo {pages} {131} (\bibinfo {year}
  {2001})}\BibitemShut {NoStop}%
\bibitem [{\citenamefont {Lutchyn}\ \emph {et~al.}(2010)\citenamefont
  {Lutchyn}, \citenamefont {Sau},\ and\ \citenamefont
  {Das~Sarma}}]{Lutchyn2010}%
  \BibitemOpen
  \bibfield  {author} {\bibinfo {author} {\bibfnamefont {R.~M.}\ \bibnamefont
  {Lutchyn}}, \bibinfo {author} {\bibfnamefont {J.~D.}\ \bibnamefont {Sau}},\
  and\ \bibinfo {author} {\bibfnamefont {S.}~\bibnamefont {Das~Sarma}},\
  }\bibfield  {title} {\bibinfo {title} {Majorana fermions and a topological
  phase transition in semiconductor-superconductor heterostructures},\ }\href
  {https://doi.org/10.1103/PhysRevLett.105.077001} {\bibfield  {journal}
  {\bibinfo  {journal} {Phys. Rev. Lett.}\ }\textbf {\bibinfo {volume} {105}},\
  \bibinfo {pages} {077001} (\bibinfo {year} {2010})}\BibitemShut {NoStop}%
\bibitem [{\citenamefont {Oreg}\ \emph {et~al.}(2010)\citenamefont {Oreg},
  \citenamefont {Refael},\ and\ \citenamefont {von Oppen}}]{Oreg2010}%
  \BibitemOpen
  \bibfield  {author} {\bibinfo {author} {\bibfnamefont {Y.}~\bibnamefont
  {Oreg}}, \bibinfo {author} {\bibfnamefont {G.}~\bibnamefont {Refael}},\ and\
  \bibinfo {author} {\bibfnamefont {F.}~\bibnamefont {von Oppen}},\ }\bibfield
  {title} {\bibinfo {title} {Helical liquids and majorana bound states in
  quantum wires},\ }\href {https://doi.org/10.1103/PhysRevLett.105.177002}
  {\bibfield  {journal} {\bibinfo  {journal} {Phys. Rev. Lett.}\ }\textbf
  {\bibinfo {volume} {105}},\ \bibinfo {pages} {177002} (\bibinfo {year}
  {2010})}\BibitemShut {NoStop}%
\bibitem [{\citenamefont {Mourik}\ \emph {et~al.}(2012)\citenamefont {Mourik}
  \emph {et~al.}}]{Mourik}%
  \BibitemOpen
  \bibfield  {author} {\bibinfo {author} {\bibfnamefont {V.}~\bibnamefont
  {Mourik}} \emph {et~al.},\ }\bibfield  {title} {\bibinfo {title} {Signatures
  of majorana fermions in hybrid superconductor-semiconductor nanowire
  devices},\ }\href@noop {} {\bibfield  {journal} {\bibinfo  {journal}
  {Science}\ }\textbf {\bibinfo {volume} {336}},\ \bibinfo {pages} {1003}
  (\bibinfo {year} {2012})}\BibitemShut {NoStop}%
\bibitem [{\citenamefont {Deng}\ \emph {et~al.}(2016)\citenamefont {Deng} \emph
  {et~al.}}]{Deng2016}%
  \BibitemOpen
  \bibfield  {author} {\bibinfo {author} {\bibfnamefont {M.}~\bibnamefont
  {Deng}} \emph {et~al.},\ }\bibfield  {title} {\bibinfo {title} {Majorana
  bound state in a coupled quantum-dot hybrid-nanowire system},\ }\href@noop {}
  {\bibfield  {journal} {\bibinfo  {journal} {Science}\ }\textbf {\bibinfo
  {volume} {354}},\ \bibinfo {pages} {1557} (\bibinfo {year}
  {2016})}\BibitemShut {NoStop}%
\bibitem [{\citenamefont {G{\"u}l}\ \emph {et~al.}(2018)\citenamefont {G{\"u}l}
  \emph {et~al.}}]{Gul2018}%
  \BibitemOpen
  \bibfield  {author} {\bibinfo {author} {\bibfnamefont {{\"O}.}~\bibnamefont
  {G{\"u}l}} \emph {et~al.},\ }\bibfield  {title} {\bibinfo {title} {Ballistic
  majorana nanowire devices},\ }\href@noop {} {\bibfield  {journal} {\bibinfo
  {journal} {Nature Nanotechnology}\ }\textbf {\bibinfo {volume} {13}},\
  \bibinfo {pages} {192} (\bibinfo {year} {2018})}\BibitemShut {NoStop}%
\bibitem [{\citenamefont {Zhang}\ \emph {et~al.}(2021)\citenamefont {Zhang}
  \emph {et~al.}}]{Zhang2021}%
  \BibitemOpen
  \bibfield  {author} {\bibinfo {author} {\bibfnamefont {H.}~\bibnamefont
  {Zhang}} \emph {et~al.},\ }\bibfield  {title} {\bibinfo {title} {Large
  zero-bias peaks in insb-al hybrid semiconductor-superconductor nanowire
  devices},\ }\href@noop {} {\bibfield  {journal} {\bibinfo  {journal} {arXiv:
  2101.11456}\ } (\bibinfo {year} {2021})}\BibitemShut {NoStop}%
\bibitem [{\citenamefont {Song}\ \emph {et~al.}(2022)\citenamefont {Song} \emph
  {et~al.}}]{Song2022}%
  \BibitemOpen
  \bibfield  {author} {\bibinfo {author} {\bibfnamefont {H.}~\bibnamefont
  {Song}} \emph {et~al.},\ }\bibfield  {title} {\bibinfo {title} {Large zero
  bias peaks and dips in a four-terminal thin inas-al nanowire device},\ }\href
  {https://doi.org/10.1103/PhysRevResearch.4.033235} {\bibfield  {journal}
  {\bibinfo  {journal} {Phys. Rev. Research}\ }\textbf {\bibinfo {volume}
  {4}},\ \bibinfo {pages} {033235} (\bibinfo {year} {2022})}\BibitemShut
  {NoStop}%
\bibitem [{\citenamefont {Wang}\ \emph {et~al.}(2022)\citenamefont {Wang} \emph
  {et~al.}}]{WangZhaoyu}%
  \BibitemOpen
  \bibfield  {author} {\bibinfo {author} {\bibfnamefont {Z.}~\bibnamefont
  {Wang}} \emph {et~al.},\ }\bibfield  {title} {\bibinfo {title} {Plateau
  regions for zero-bias peaks within 5$\%$ of the quantized conductance value
  $2{e}^{2}/h$},\ }\href {https://doi.org/10.1103/PhysRevLett.129.167702}
  {\bibfield  {journal} {\bibinfo  {journal} {Phys. Rev. Lett.}\ }\textbf
  {\bibinfo {volume} {129}},\ \bibinfo {pages} {167702} (\bibinfo {year}
  {2022})}\BibitemShut {NoStop}%
\bibitem [{\citenamefont {Prada}\ \emph {et~al.}(2020)\citenamefont {Prada}
  \emph {et~al.}}]{Prada2020}%
  \BibitemOpen
  \bibfield  {author} {\bibinfo {author} {\bibfnamefont {E.}~\bibnamefont
  {Prada}} \emph {et~al.},\ }\bibfield  {title} {\bibinfo {title} {From andreev
  to majorana bound states in hybrid superconductor--semiconductor nanowires},\
  }\href@noop {} {\bibfield  {journal} {\bibinfo  {journal} {Nature Reviews
  Physics}\ }\textbf {\bibinfo {volume} {2}},\ \bibinfo {pages} {575} (\bibinfo
  {year} {2020})}\BibitemShut {NoStop}%
\bibitem [{\citenamefont {Zhang}\ \emph {et~al.}(2019)\citenamefont {Zhang}
  \emph {et~al.}}]{NextSteps}%
  \BibitemOpen
  \bibfield  {author} {\bibinfo {author} {\bibfnamefont {H.}~\bibnamefont
  {Zhang}} \emph {et~al.},\ }\bibfield  {title} {\bibinfo {title} {Next steps
  of quantum transport in majorana nanowire devices},\ }\href@noop {}
  {\bibfield  {journal} {\bibinfo  {journal} {Nature Communications}\ }\textbf
  {\bibinfo {volume} {10}},\ \bibinfo {pages} {5128} (\bibinfo {year}
  {2019})}\BibitemShut {NoStop}%
\bibitem [{\citenamefont {Alicea}\ \emph {et~al.}(2010)\citenamefont {Alicea},
  \citenamefont {Oreg}, \citenamefont {Refael}, \citenamefont {Oppen},\ and\
  \citenamefont {Fisher}}]{Alicea_NP}%
  \BibitemOpen
  \bibfield  {author} {\bibinfo {author} {\bibfnamefont {J.}~\bibnamefont
  {Alicea}}, \bibinfo {author} {\bibfnamefont {Y.}~\bibnamefont {Oreg}},
  \bibinfo {author} {\bibfnamefont {G.}~\bibnamefont {Refael}}, \bibinfo
  {author} {\bibfnamefont {F.}~\bibnamefont {Oppen}},\ and\ \bibinfo {author}
  {\bibfnamefont {M.}~\bibnamefont {Fisher}},\ }\bibfield  {title} {\bibinfo
  {title} {Non-abelian statistics and topological quantum information
  processing in 1d wire networks},\ }\href {https://doi.org/10.1038/nphys1915}
  {\bibfield  {journal} {\bibinfo  {journal} {Nature Physics}\ }\textbf
  {\bibinfo {volume} {7}} (\bibinfo {year} {2010})}\BibitemShut {NoStop}%
\bibitem [{\citenamefont {Hyart}\ \emph {et~al.}(2013)\citenamefont {Hyart},
  \citenamefont {van Heck}, \citenamefont {Fulga}, \citenamefont {Burrello},
  \citenamefont {Akhmerov},\ and\ \citenamefont {Beenakker}}]{Hyart_PRB}%
  \BibitemOpen
  \bibfield  {author} {\bibinfo {author} {\bibfnamefont {T.}~\bibnamefont
  {Hyart}}, \bibinfo {author} {\bibfnamefont {B.}~\bibnamefont {van Heck}},
  \bibinfo {author} {\bibfnamefont {I.~C.}\ \bibnamefont {Fulga}}, \bibinfo
  {author} {\bibfnamefont {M.}~\bibnamefont {Burrello}}, \bibinfo {author}
  {\bibfnamefont {A.~R.}\ \bibnamefont {Akhmerov}},\ and\ \bibinfo {author}
  {\bibfnamefont {C.~W.~J.}\ \bibnamefont {Beenakker}},\ }\bibfield  {title}
  {\bibinfo {title} {Flux-controlled quantum computation with majorana
  fermions},\ }\href {https://doi.org/10.1103/PhysRevB.88.035121} {\bibfield
  {journal} {\bibinfo  {journal} {Phys. Rev. B}\ }\textbf {\bibinfo {volume}
  {88}},\ \bibinfo {pages} {035121} (\bibinfo {year} {2013})}\BibitemShut
  {NoStop}%
\bibitem [{\citenamefont {Knapp}\ \emph {et~al.}(2016)\citenamefont {Knapp},
  \citenamefont {Zaletel}, \citenamefont {Liu}, \citenamefont {Cheng},
  \citenamefont {Bonderson},\ and\ \citenamefont {Nayak}}]{Dong_PRX_2016}%
  \BibitemOpen
  \bibfield  {author} {\bibinfo {author} {\bibfnamefont {C.}~\bibnamefont
  {Knapp}}, \bibinfo {author} {\bibfnamefont {M.}~\bibnamefont {Zaletel}},
  \bibinfo {author} {\bibfnamefont {D.~E.}\ \bibnamefont {Liu}}, \bibinfo
  {author} {\bibfnamefont {M.}~\bibnamefont {Cheng}}, \bibinfo {author}
  {\bibfnamefont {P.}~\bibnamefont {Bonderson}},\ and\ \bibinfo {author}
  {\bibfnamefont {C.}~\bibnamefont {Nayak}},\ }\bibfield  {title} {\bibinfo
  {title} {The nature and correction of diabatic errors in anyon braiding},\
  }\href {https://doi.org/10.1103/PhysRevX.6.041003} {\bibfield  {journal}
  {\bibinfo  {journal} {Phys. Rev. X}\ }\textbf {\bibinfo {volume} {6}},\
  \bibinfo {pages} {041003} (\bibinfo {year} {2016})}\BibitemShut {NoStop}%
\bibitem [{\citenamefont {Aasen}\ \emph {et~al.}(2016)\citenamefont {Aasen},
  \citenamefont {Hell}, \citenamefont {Mishmash}, \citenamefont {Higginbotham},
  \citenamefont {Danon}, \citenamefont {Leijnse}, \citenamefont {Jespersen},
  \citenamefont {Folk}, \citenamefont {Marcus}, \citenamefont {Flensberg},\
  and\ \citenamefont {Alicea}}]{Alicea_PRX}%
  \BibitemOpen
  \bibfield  {author} {\bibinfo {author} {\bibfnamefont {D.}~\bibnamefont
  {Aasen}}, \bibinfo {author} {\bibfnamefont {M.}~\bibnamefont {Hell}},
  \bibinfo {author} {\bibfnamefont {R.~V.}\ \bibnamefont {Mishmash}}, \bibinfo
  {author} {\bibfnamefont {A.}~\bibnamefont {Higginbotham}}, \bibinfo {author}
  {\bibfnamefont {J.}~\bibnamefont {Danon}}, \bibinfo {author} {\bibfnamefont
  {M.}~\bibnamefont {Leijnse}}, \bibinfo {author} {\bibfnamefont {T.~S.}\
  \bibnamefont {Jespersen}}, \bibinfo {author} {\bibfnamefont {J.~A.}\
  \bibnamefont {Folk}}, \bibinfo {author} {\bibfnamefont {C.~M.}\ \bibnamefont
  {Marcus}}, \bibinfo {author} {\bibfnamefont {K.}~\bibnamefont {Flensberg}},\
  and\ \bibinfo {author} {\bibfnamefont {J.}~\bibnamefont {Alicea}},\
  }\bibfield  {title} {\bibinfo {title} {Milestones toward majorana-based
  quantum computing},\ }\href {https://doi.org/10.1103/PhysRevX.6.031016}
  {\bibfield  {journal} {\bibinfo  {journal} {Phys. Rev. X}\ }\textbf {\bibinfo
  {volume} {6}},\ \bibinfo {pages} {031016} (\bibinfo {year}
  {2016})}\BibitemShut {NoStop}%
\bibitem [{\citenamefont {Plugge}\ \emph {et~al.}(2017)\citenamefont {Plugge},
  \citenamefont {Rasmussen}, \citenamefont {Egger},\ and\ \citenamefont
  {Flensberg}}]{Plugge_2017}%
  \BibitemOpen
  \bibfield  {author} {\bibinfo {author} {\bibfnamefont {S.}~\bibnamefont
  {Plugge}}, \bibinfo {author} {\bibfnamefont {A.}~\bibnamefont {Rasmussen}},
  \bibinfo {author} {\bibfnamefont {R.}~\bibnamefont {Egger}},\ and\ \bibinfo
  {author} {\bibfnamefont {K.}~\bibnamefont {Flensberg}},\ }\bibfield  {title}
  {\bibinfo {title} {Majorana box qubits},\ }\href
  {https://doi.org/10.1088/1367-2630/aa54e1} {\bibfield  {journal} {\bibinfo
  {journal} {New Journal of Physics}\ }\textbf {\bibinfo {volume} {19}},\
  \bibinfo {pages} {012001} (\bibinfo {year} {2017})}\BibitemShut {NoStop}%
\bibitem [{\citenamefont {Vijay}\ and\ \citenamefont {Fu}(2016)}]{Fu_braiding}%
  \BibitemOpen
  \bibfield  {author} {\bibinfo {author} {\bibfnamefont {S.}~\bibnamefont
  {Vijay}}\ and\ \bibinfo {author} {\bibfnamefont {L.}~\bibnamefont {Fu}},\
  }\bibfield  {title} {\bibinfo {title} {Teleportation-based quantum
  information processing with majorana zero modes},\ }\href
  {https://doi.org/10.1103/PhysRevB.94.235446} {\bibfield  {journal} {\bibinfo
  {journal} {Phys. Rev. B}\ }\textbf {\bibinfo {volume} {94}},\ \bibinfo
  {pages} {235446} (\bibinfo {year} {2016})}\BibitemShut {NoStop}%
\bibitem [{\citenamefont {Karzig}\ \emph {et~al.}(2017)\citenamefont {Karzig},
  \citenamefont {Knapp}, \citenamefont {Lutchyn}, \citenamefont {Bonderson},
  \citenamefont {Hastings}, \citenamefont {Nayak}, \citenamefont {Alicea},
  \citenamefont {Flensberg}, \citenamefont {Plugge}, \citenamefont {Oreg},
  \citenamefont {Marcus},\ and\ \citenamefont {Freedman}}]{MS_scalable}%
  \BibitemOpen
  \bibfield  {author} {\bibinfo {author} {\bibfnamefont {T.}~\bibnamefont
  {Karzig}}, \bibinfo {author} {\bibfnamefont {C.}~\bibnamefont {Knapp}},
  \bibinfo {author} {\bibfnamefont {R.~M.}\ \bibnamefont {Lutchyn}}, \bibinfo
  {author} {\bibfnamefont {P.}~\bibnamefont {Bonderson}}, \bibinfo {author}
  {\bibfnamefont {M.~B.}\ \bibnamefont {Hastings}}, \bibinfo {author}
  {\bibfnamefont {C.}~\bibnamefont {Nayak}}, \bibinfo {author} {\bibfnamefont
  {J.}~\bibnamefont {Alicea}}, \bibinfo {author} {\bibfnamefont
  {K.}~\bibnamefont {Flensberg}}, \bibinfo {author} {\bibfnamefont
  {S.}~\bibnamefont {Plugge}}, \bibinfo {author} {\bibfnamefont
  {Y.}~\bibnamefont {Oreg}}, \bibinfo {author} {\bibfnamefont {C.~M.}\
  \bibnamefont {Marcus}},\ and\ \bibinfo {author} {\bibfnamefont {M.~H.}\
  \bibnamefont {Freedman}},\ }\bibfield  {title} {\bibinfo {title} {Scalable
  designs for quasiparticle-poisoning-protected topological quantum computation
  with majorana zero modes},\ }\href
  {https://doi.org/10.1103/PhysRevB.95.235305} {\bibfield  {journal} {\bibinfo
  {journal} {Phys. Rev. B}\ }\textbf {\bibinfo {volume} {95}},\ \bibinfo
  {pages} {235305} (\bibinfo {year} {2017})}\BibitemShut {NoStop}%
\bibitem [{\citenamefont {Koch}\ \emph {et~al.}(2007)\citenamefont {Koch},
  \citenamefont {Yu}, \citenamefont {Gambetta}, \citenamefont {Houck},
  \citenamefont {Schuster}, \citenamefont {Majer}, \citenamefont {Blais},
  \citenamefont {Devoret}, \citenamefont {Girvin},\ and\ \citenamefont
  {Schoelkopf}}]{Transmon}%
  \BibitemOpen
  \bibfield  {author} {\bibinfo {author} {\bibfnamefont {J.}~\bibnamefont
  {Koch}}, \bibinfo {author} {\bibfnamefont {T.~M.}\ \bibnamefont {Yu}},
  \bibinfo {author} {\bibfnamefont {J.}~\bibnamefont {Gambetta}}, \bibinfo
  {author} {\bibfnamefont {A.~A.}\ \bibnamefont {Houck}}, \bibinfo {author}
  {\bibfnamefont {D.~I.}\ \bibnamefont {Schuster}}, \bibinfo {author}
  {\bibfnamefont {J.}~\bibnamefont {Majer}}, \bibinfo {author} {\bibfnamefont
  {A.}~\bibnamefont {Blais}}, \bibinfo {author} {\bibfnamefont {M.~H.}\
  \bibnamefont {Devoret}}, \bibinfo {author} {\bibfnamefont {S.~M.}\
  \bibnamefont {Girvin}},\ and\ \bibinfo {author} {\bibfnamefont {R.~J.}\
  \bibnamefont {Schoelkopf}},\ }\bibfield  {title} {\bibinfo {title}
  {Charge-insensitive qubit design derived from the cooper pair box},\ }\href
  {https://doi.org/10.1103/PhysRevA.76.042319} {\bibfield  {journal} {\bibinfo
  {journal} {Phys. Rev. A}\ }\textbf {\bibinfo {volume} {76}},\ \bibinfo
  {pages} {042319} (\bibinfo {year} {2007})}\BibitemShut {NoStop}%
\bibitem [{\citenamefont {Houck}\ \emph {et~al.}(2007)\citenamefont {Houck},
  \citenamefont {Schuster}, \citenamefont {Gambetta}, \citenamefont {Schreier},
  \citenamefont {Johnson}, \citenamefont {Chow}, \citenamefont {Frunzio},
  \citenamefont {Majer}, \citenamefont {Devoret}, \citenamefont {Girvin},\ and\
  \citenamefont {Schoelkopf}}]{Transmon_Nature}%
  \BibitemOpen
  \bibfield  {author} {\bibinfo {author} {\bibfnamefont {A.}~\bibnamefont
  {Houck}}, \bibinfo {author} {\bibfnamefont {D.}~\bibnamefont {Schuster}},
  \bibinfo {author} {\bibfnamefont {J.}~\bibnamefont {Gambetta}}, \bibinfo
  {author} {\bibfnamefont {J.}~\bibnamefont {Schreier}}, \bibinfo {author}
  {\bibfnamefont {B.}~\bibnamefont {Johnson}}, \bibinfo {author} {\bibfnamefont
  {J.}~\bibnamefont {Chow}}, \bibinfo {author} {\bibfnamefont {L.}~\bibnamefont
  {Frunzio}}, \bibinfo {author} {\bibfnamefont {J.}~\bibnamefont {Majer}},
  \bibinfo {author} {\bibfnamefont {M.}~\bibnamefont {Devoret}}, \bibinfo
  {author} {\bibfnamefont {S.}~\bibnamefont {Girvin}},\ and\ \bibinfo {author}
  {\bibfnamefont {R.}~\bibnamefont {Schoelkopf}},\ }\bibfield  {title}
  {\bibinfo {title} {Generating single microwave photons in a circuit},\ }\href
  {https://doi.org/10.1038/nature06126} {\bibfield  {journal} {\bibinfo
  {journal} {Nature}\ }\textbf {\bibinfo {volume} {449}},\ \bibinfo {pages}
  {328} (\bibinfo {year} {2007})}\BibitemShut {NoStop}%
\bibitem [{\citenamefont {Ginossar}\ and\ \citenamefont
  {Grosfeld}(2014)}]{Majorana_transmon}%
  \BibitemOpen
  \bibfield  {author} {\bibinfo {author} {\bibfnamefont {E.}~\bibnamefont
  {Ginossar}}\ and\ \bibinfo {author} {\bibfnamefont {E.}~\bibnamefont
  {Grosfeld}},\ }\bibfield  {title} {\bibinfo {title} {Microwave transitions as
  a signature of coherent parity mixing effects in the majorana-transmon
  qubit},\ }\href {https://doi.org/10.1038/ncomms5772} {\bibfield  {journal}
  {\bibinfo  {journal} {Nature communications}\ }\textbf {\bibinfo {volume}
  {5}},\ \bibinfo {pages} {4772} (\bibinfo {year} {2014})}\BibitemShut
  {NoStop}%
\bibitem [{\citenamefont {Yavilberg}\ \emph {et~al.}(2015)\citenamefont
  {Yavilberg}, \citenamefont {Ginossar},\ and\ \citenamefont
  {Grosfeld}}]{2015_PRB_MZM_transmon}%
  \BibitemOpen
  \bibfield  {author} {\bibinfo {author} {\bibfnamefont {K.}~\bibnamefont
  {Yavilberg}}, \bibinfo {author} {\bibfnamefont {E.}~\bibnamefont
  {Ginossar}},\ and\ \bibinfo {author} {\bibfnamefont {E.}~\bibnamefont
  {Grosfeld}},\ }\bibfield  {title} {\bibinfo {title} {Fermion parity
  measurement and control in majorana circuit quantum electrodynamics},\ }\href
  {https://doi.org/10.1103/PhysRevB.92.075143} {\bibfield  {journal} {\bibinfo
  {journal} {Phys. Rev. B}\ }\textbf {\bibinfo {volume} {92}},\ \bibinfo
  {pages} {075143} (\bibinfo {year} {2015})}\BibitemShut {NoStop}%
\bibitem [{\citenamefont {Li}\ \emph {et~al.}(2018)\citenamefont {Li},
  \citenamefont {Coish}, \citenamefont {Hell}, \citenamefont {Flensberg},\ and\
  \citenamefont {Leijnse}}]{2018_PRB_MZM_transmon}%
  \BibitemOpen
  \bibfield  {author} {\bibinfo {author} {\bibfnamefont {T.}~\bibnamefont
  {Li}}, \bibinfo {author} {\bibfnamefont {W.~A.}\ \bibnamefont {Coish}},
  \bibinfo {author} {\bibfnamefont {M.}~\bibnamefont {Hell}}, \bibinfo {author}
  {\bibfnamefont {K.}~\bibnamefont {Flensberg}},\ and\ \bibinfo {author}
  {\bibfnamefont {M.}~\bibnamefont {Leijnse}},\ }\bibfield  {title} {\bibinfo
  {title} {Four-majorana qubit with charge readout: Dynamics and decoherence},\
  }\href {https://doi.org/10.1103/PhysRevB.98.205403} {\bibfield  {journal}
  {\bibinfo  {journal} {Phys. Rev. B}\ }\textbf {\bibinfo {volume} {98}},\
  \bibinfo {pages} {205403} (\bibinfo {year} {2018})}\BibitemShut {NoStop}%
\bibitem [{\citenamefont {\'Avila}\ \emph {et~al.}(2020)\citenamefont
  {\'Avila}, \citenamefont {Prada}, \citenamefont {San-Jose},\ and\
  \citenamefont {Aguado}}]{2020_PRR_MZM_transmon}%
  \BibitemOpen
  \bibfield  {author} {\bibinfo {author} {\bibfnamefont {J.}~\bibnamefont
  {\'Avila}}, \bibinfo {author} {\bibfnamefont {E.}~\bibnamefont {Prada}},
  \bibinfo {author} {\bibfnamefont {P.}~\bibnamefont {San-Jose}},\ and\
  \bibinfo {author} {\bibfnamefont {R.}~\bibnamefont {Aguado}},\ }\bibfield
  {title} {\bibinfo {title} {Majorana oscillations and parity crossings in
  semiconductor nanowire-based transmon qubits},\ }\href
  {https://doi.org/10.1103/PhysRevResearch.2.033493} {\bibfield  {journal}
  {\bibinfo  {journal} {Phys. Rev. Research}\ }\textbf {\bibinfo {volume}
  {2}},\ \bibinfo {pages} {033493} (\bibinfo {year} {2020})}\BibitemShut
  {NoStop}%
\bibitem [{\citenamefont {Chirolli}\ \emph {et~al.}(2022)\citenamefont
  {Chirolli}, \citenamefont {Yao},\ and\ \citenamefont
  {Moore}}]{2022_PRL_MZM_transmon}%
  \BibitemOpen
  \bibfield  {author} {\bibinfo {author} {\bibfnamefont {L.}~\bibnamefont
  {Chirolli}}, \bibinfo {author} {\bibfnamefont {N.~Y.}\ \bibnamefont {Yao}},\
  and\ \bibinfo {author} {\bibfnamefont {J.~E.}\ \bibnamefont {Moore}},\
  }\bibfield  {title} {\bibinfo {title} {Swap gate between a majorana qubit and
  a parity-protected superconducting qubit},\ }\href
  {https://doi.org/10.1103/PhysRevLett.129.177701} {\bibfield  {journal}
  {\bibinfo  {journal} {Phys. Rev. Lett.}\ }\textbf {\bibinfo {volume} {129}},\
  \bibinfo {pages} {177701} (\bibinfo {year} {2022})}\BibitemShut {NoStop}%
\bibitem [{\citenamefont {Larsen}\ \emph {et~al.}(2015)\citenamefont {Larsen},
  \citenamefont {Petersson}, \citenamefont {Kuemmeth}, \citenamefont
  {Jespersen}, \citenamefont {Krogstrup}, \citenamefont {Nyg\aa{}rd},\ and\
  \citenamefont {Marcus}}]{2015_PRL_gatemon}%
  \BibitemOpen
  \bibfield  {author} {\bibinfo {author} {\bibfnamefont {T.~W.}\ \bibnamefont
  {Larsen}}, \bibinfo {author} {\bibfnamefont {K.~D.}\ \bibnamefont
  {Petersson}}, \bibinfo {author} {\bibfnamefont {F.}~\bibnamefont {Kuemmeth}},
  \bibinfo {author} {\bibfnamefont {T.~S.}\ \bibnamefont {Jespersen}}, \bibinfo
  {author} {\bibfnamefont {P.}~\bibnamefont {Krogstrup}}, \bibinfo {author}
  {\bibfnamefont {J.}~\bibnamefont {Nyg\aa{}rd}},\ and\ \bibinfo {author}
  {\bibfnamefont {C.~M.}\ \bibnamefont {Marcus}},\ }\bibfield  {title}
  {\bibinfo {title} {Semiconductor-nanowire-based superconducting qubit},\
  }\href {https://doi.org/10.1103/PhysRevLett.115.127001} {\bibfield  {journal}
  {\bibinfo  {journal} {Phys. Rev. Lett.}\ }\textbf {\bibinfo {volume} {115}},\
  \bibinfo {pages} {127001} (\bibinfo {year} {2015})}\BibitemShut {NoStop}%
\bibitem [{\citenamefont {de~Lange}\ \emph {et~al.}(2015)\citenamefont
  {de~Lange}, \citenamefont {van Heck}, \citenamefont {Bruno}, \citenamefont
  {van Woerkom}, \citenamefont {Geresdi}, \citenamefont {Plissard},
  \citenamefont {Bakkers}, \citenamefont {Akhmerov},\ and\ \citenamefont
  {DiCarlo}}]{DiCarlo_2015}%
  \BibitemOpen
  \bibfield  {author} {\bibinfo {author} {\bibfnamefont {G.}~\bibnamefont
  {de~Lange}}, \bibinfo {author} {\bibfnamefont {B.}~\bibnamefont {van Heck}},
  \bibinfo {author} {\bibfnamefont {A.}~\bibnamefont {Bruno}}, \bibinfo
  {author} {\bibfnamefont {D.~J.}\ \bibnamefont {van Woerkom}}, \bibinfo
  {author} {\bibfnamefont {A.}~\bibnamefont {Geresdi}}, \bibinfo {author}
  {\bibfnamefont {S.~R.}\ \bibnamefont {Plissard}}, \bibinfo {author}
  {\bibfnamefont {E.~P. A.~M.}\ \bibnamefont {Bakkers}}, \bibinfo {author}
  {\bibfnamefont {A.~R.}\ \bibnamefont {Akhmerov}},\ and\ \bibinfo {author}
  {\bibfnamefont {L.}~\bibnamefont {DiCarlo}},\ }\bibfield  {title} {\bibinfo
  {title} {Realization of microwave quantum circuits using hybrid
  superconducting-semiconducting nanowire josephson elements},\ }\href
  {https://doi.org/10.1103/PhysRevLett.115.127002} {\bibfield  {journal}
  {\bibinfo  {journal} {Phys. Rev. Lett.}\ }\textbf {\bibinfo {volume} {115}},\
  \bibinfo {pages} {127002} (\bibinfo {year} {2015})}\BibitemShut {NoStop}%
\bibitem [{\citenamefont {Casparis}\ \emph {et~al.}(2016)\citenamefont
  {Casparis}, \citenamefont {Larsen}, \citenamefont {Olsen}, \citenamefont
  {Kuemmeth}, \citenamefont {Krogstrup}, \citenamefont {Nyg\aa{}rd},
  \citenamefont {Petersson},\ and\ \citenamefont {Marcus}}]{Gatemon_2016_PRL}%
  \BibitemOpen
  \bibfield  {author} {\bibinfo {author} {\bibfnamefont {L.}~\bibnamefont
  {Casparis}}, \bibinfo {author} {\bibfnamefont {T.~W.}\ \bibnamefont
  {Larsen}}, \bibinfo {author} {\bibfnamefont {M.~S.}\ \bibnamefont {Olsen}},
  \bibinfo {author} {\bibfnamefont {F.}~\bibnamefont {Kuemmeth}}, \bibinfo
  {author} {\bibfnamefont {P.}~\bibnamefont {Krogstrup}}, \bibinfo {author}
  {\bibfnamefont {J.}~\bibnamefont {Nyg\aa{}rd}}, \bibinfo {author}
  {\bibfnamefont {K.~D.}\ \bibnamefont {Petersson}},\ and\ \bibinfo {author}
  {\bibfnamefont {C.~M.}\ \bibnamefont {Marcus}},\ }\bibfield  {title}
  {\bibinfo {title} {Gatemon benchmarking and two-qubit operations},\ }\href
  {https://doi.org/10.1103/PhysRevLett.116.150505} {\bibfield  {journal}
  {\bibinfo  {journal} {Phys. Rev. Lett.}\ }\textbf {\bibinfo {volume} {116}},\
  \bibinfo {pages} {150505} (\bibinfo {year} {2016})}\BibitemShut {NoStop}%
\bibitem [{\citenamefont {Luthi}\ \emph {et~al.}(2018)\citenamefont {Luthi},
  \citenamefont {Stavenga}, \citenamefont {Enzing}, \citenamefont {Bruno},
  \citenamefont {Dickel}, \citenamefont {Langford}, \citenamefont {Rol},
  \citenamefont {Jespersen}, \citenamefont {Nyg\aa{}rd}, \citenamefont
  {Krogstrup},\ and\ \citenamefont {DiCarlo}}]{DiCarlo_2018}%
  \BibitemOpen
  \bibfield  {author} {\bibinfo {author} {\bibfnamefont {F.}~\bibnamefont
  {Luthi}}, \bibinfo {author} {\bibfnamefont {T.}~\bibnamefont {Stavenga}},
  \bibinfo {author} {\bibfnamefont {O.~W.}\ \bibnamefont {Enzing}}, \bibinfo
  {author} {\bibfnamefont {A.}~\bibnamefont {Bruno}}, \bibinfo {author}
  {\bibfnamefont {C.}~\bibnamefont {Dickel}}, \bibinfo {author} {\bibfnamefont
  {N.~K.}\ \bibnamefont {Langford}}, \bibinfo {author} {\bibfnamefont {M.~A.}\
  \bibnamefont {Rol}}, \bibinfo {author} {\bibfnamefont {T.~S.}\ \bibnamefont
  {Jespersen}}, \bibinfo {author} {\bibfnamefont {J.}~\bibnamefont
  {Nyg\aa{}rd}}, \bibinfo {author} {\bibfnamefont {P.}~\bibnamefont
  {Krogstrup}},\ and\ \bibinfo {author} {\bibfnamefont {L.}~\bibnamefont
  {DiCarlo}},\ }\bibfield  {title} {\bibinfo {title} {Evolution of nanowire
  transmon qubits and their coherence in a magnetic field},\ }\href
  {https://doi.org/10.1103/PhysRevLett.120.100502} {\bibfield  {journal}
  {\bibinfo  {journal} {Phys. Rev. Lett.}\ }\textbf {\bibinfo {volume} {120}},\
  \bibinfo {pages} {100502} (\bibinfo {year} {2018})}\BibitemShut {NoStop}%
\bibitem [{\citenamefont {Kringh\o{}j}\ \emph {et~al.}(2020)\citenamefont
  {Kringh\o{}j}, \citenamefont {van Heck}, \citenamefont {Larsen},
  \citenamefont {Erlandsson}, \citenamefont {Sabonis}, \citenamefont
  {Krogstrup}, \citenamefont {Casparis}, \citenamefont {Petersson},\ and\
  \citenamefont {Marcus}}]{2020_PRL_Charlie_transmon}%
  \BibitemOpen
  \bibfield  {author} {\bibinfo {author} {\bibfnamefont {A.}~\bibnamefont
  {Kringh\o{}j}}, \bibinfo {author} {\bibfnamefont {B.}~\bibnamefont {van
  Heck}}, \bibinfo {author} {\bibfnamefont {T.~W.}\ \bibnamefont {Larsen}},
  \bibinfo {author} {\bibfnamefont {O.}~\bibnamefont {Erlandsson}}, \bibinfo
  {author} {\bibfnamefont {D.}~\bibnamefont {Sabonis}}, \bibinfo {author}
  {\bibfnamefont {P.}~\bibnamefont {Krogstrup}}, \bibinfo {author}
  {\bibfnamefont {L.}~\bibnamefont {Casparis}}, \bibinfo {author}
  {\bibfnamefont {K.~D.}\ \bibnamefont {Petersson}},\ and\ \bibinfo {author}
  {\bibfnamefont {C.~M.}\ \bibnamefont {Marcus}},\ }\bibfield  {title}
  {\bibinfo {title} {Suppressed charge dispersion via resonant tunneling in a
  single-channel transmon},\ }\href
  {https://doi.org/10.1103/PhysRevLett.124.246803} {\bibfield  {journal}
  {\bibinfo  {journal} {Phys. Rev. Lett.}\ }\textbf {\bibinfo {volume} {124}},\
  \bibinfo {pages} {246803} (\bibinfo {year} {2020})}\BibitemShut {NoStop}%
\bibitem [{\citenamefont {Bargerbos}\ \emph {et~al.}(2020)\citenamefont
  {Bargerbos}, \citenamefont {Uilhoorn}, \citenamefont {Yang}, \citenamefont
  {Krogstrup}, \citenamefont {Kouwenhoven}, \citenamefont {de~Lange},
  \citenamefont {van Heck},\ and\ \citenamefont {Kou}}]{2020_PRL_Leo_transmon}%
  \BibitemOpen
  \bibfield  {author} {\bibinfo {author} {\bibfnamefont {A.}~\bibnamefont
  {Bargerbos}}, \bibinfo {author} {\bibfnamefont {W.}~\bibnamefont {Uilhoorn}},
  \bibinfo {author} {\bibfnamefont {C.-K.}\ \bibnamefont {Yang}}, \bibinfo
  {author} {\bibfnamefont {P.}~\bibnamefont {Krogstrup}}, \bibinfo {author}
  {\bibfnamefont {L.~P.}\ \bibnamefont {Kouwenhoven}}, \bibinfo {author}
  {\bibfnamefont {G.}~\bibnamefont {de~Lange}}, \bibinfo {author}
  {\bibfnamefont {B.}~\bibnamefont {van Heck}},\ and\ \bibinfo {author}
  {\bibfnamefont {A.}~\bibnamefont {Kou}},\ }\bibfield  {title} {\bibinfo
  {title} {Observation of vanishing charge dispersion of a nearly open
  superconducting island},\ }\href
  {https://doi.org/10.1103/PhysRevLett.124.246802} {\bibfield  {journal}
  {\bibinfo  {journal} {Phys. Rev. Lett.}\ }\textbf {\bibinfo {volume} {124}},\
  \bibinfo {pages} {246802} (\bibinfo {year} {2020})}\BibitemShut {NoStop}%
\bibitem [{\citenamefont {Larsen}\ \emph {et~al.}(2020)\citenamefont {Larsen},
  \citenamefont {Gershenson}, \citenamefont {Casparis}, \citenamefont
  {Kringh\o{}j}, \citenamefont {Pearson}, \citenamefont {McNeil}, \citenamefont
  {Kuemmeth}, \citenamefont {Krogstrup}, \citenamefont {Petersson},\ and\
  \citenamefont {Marcus}}]{2020_PRL_pi_qubit}%
  \BibitemOpen
  \bibfield  {author} {\bibinfo {author} {\bibfnamefont {T.~W.}\ \bibnamefont
  {Larsen}}, \bibinfo {author} {\bibfnamefont {M.~E.}\ \bibnamefont
  {Gershenson}}, \bibinfo {author} {\bibfnamefont {L.}~\bibnamefont
  {Casparis}}, \bibinfo {author} {\bibfnamefont {A.}~\bibnamefont
  {Kringh\o{}j}}, \bibinfo {author} {\bibfnamefont {N.~J.}\ \bibnamefont
  {Pearson}}, \bibinfo {author} {\bibfnamefont {R.~P.~G.}\ \bibnamefont
  {McNeil}}, \bibinfo {author} {\bibfnamefont {F.}~\bibnamefont {Kuemmeth}},
  \bibinfo {author} {\bibfnamefont {P.}~\bibnamefont {Krogstrup}}, \bibinfo
  {author} {\bibfnamefont {K.~D.}\ \bibnamefont {Petersson}},\ and\ \bibinfo
  {author} {\bibfnamefont {C.~M.}\ \bibnamefont {Marcus}},\ }\bibfield  {title}
  {\bibinfo {title} {Parity-protected superconductor-semiconductor qubit},\
  }\href {https://doi.org/10.1103/PhysRevLett.125.056801} {\bibfield  {journal}
  {\bibinfo  {journal} {Phys. Rev. Lett.}\ }\textbf {\bibinfo {volume} {125}},\
  \bibinfo {pages} {056801} (\bibinfo {year} {2020})}\BibitemShut {NoStop}%
\bibitem [{\citenamefont {Sabonis}\ \emph {et~al.}(2020)\citenamefont
  {Sabonis}, \citenamefont {Erlandsson}, \citenamefont {Kringh\o{}j},
  \citenamefont {van Heck}, \citenamefont {Larsen}, \citenamefont {Petkovic},
  \citenamefont {Krogstrup}, \citenamefont {Petersson},\ and\ \citenamefont
  {Marcus}}]{2020_PRL_fullshell}%
  \BibitemOpen
  \bibfield  {author} {\bibinfo {author} {\bibfnamefont {D.}~\bibnamefont
  {Sabonis}}, \bibinfo {author} {\bibfnamefont {O.}~\bibnamefont {Erlandsson}},
  \bibinfo {author} {\bibfnamefont {A.}~\bibnamefont {Kringh\o{}j}}, \bibinfo
  {author} {\bibfnamefont {B.}~\bibnamefont {van Heck}}, \bibinfo {author}
  {\bibfnamefont {T.~W.}\ \bibnamefont {Larsen}}, \bibinfo {author}
  {\bibfnamefont {I.}~\bibnamefont {Petkovic}}, \bibinfo {author}
  {\bibfnamefont {P.}~\bibnamefont {Krogstrup}}, \bibinfo {author}
  {\bibfnamefont {K.~D.}\ \bibnamefont {Petersson}},\ and\ \bibinfo {author}
  {\bibfnamefont {C.~M.}\ \bibnamefont {Marcus}},\ }\bibfield  {title}
  {\bibinfo {title} {Destructive little-parks effect in a full-shell
  nanowire-based transmon},\ }\href
  {https://doi.org/10.1103/PhysRevLett.125.156804} {\bibfield  {journal}
  {\bibinfo  {journal} {Phys. Rev. Lett.}\ }\textbf {\bibinfo {volume} {125}},\
  \bibinfo {pages} {156804} (\bibinfo {year} {2020})}\BibitemShut {NoStop}%
\bibitem [{\citenamefont {Bargerbos}\ \emph {et~al.}(2022)\citenamefont
  {Bargerbos} \emph {et~al.}}]{2022_PRXQuantum}%
  \BibitemOpen
  \bibfield  {author} {\bibinfo {author} {\bibfnamefont {A.}~\bibnamefont
  {Bargerbos}} \emph {et~al.},\ }\bibfield  {title} {\bibinfo {title}
  {Singlet-doublet transitions of a quantum dot josephson junction detected in
  a transmon circuit},\ }\href {https://doi.org/10.1103/PRXQuantum.3.030311}
  {\bibfield  {journal} {\bibinfo  {journal} {PRX Quantum}\ }\textbf {\bibinfo
  {volume} {3}},\ \bibinfo {pages} {030311} (\bibinfo {year}
  {2022})}\BibitemShut {NoStop}%
\bibitem [{\citenamefont {Caroff}\ \emph {et~al.}(2009)\citenamefont {Caroff},
  \citenamefont {Dick}, \citenamefont {Johansson}, \citenamefont {Messing},
  \citenamefont {Deppert},\ and\ \citenamefont {Samuelson}}]{Caroff2009}%
  \BibitemOpen
  \bibfield  {author} {\bibinfo {author} {\bibfnamefont {P.}~\bibnamefont
  {Caroff}}, \bibinfo {author} {\bibfnamefont {K.~A.}\ \bibnamefont {Dick}},
  \bibinfo {author} {\bibfnamefont {J.}~\bibnamefont {Johansson}}, \bibinfo
  {author} {\bibfnamefont {M.~E.}\ \bibnamefont {Messing}}, \bibinfo {author}
  {\bibfnamefont {K.}~\bibnamefont {Deppert}},\ and\ \bibinfo {author}
  {\bibfnamefont {L.}~\bibnamefont {Samuelson}},\ }\bibfield  {title} {\bibinfo
  {title} {Controlled polytypic and twin-plane superlattices in iii-v
  nanowires},\ }\href@noop {} {\bibfield  {journal} {\bibinfo  {journal}
  {Nature nanotechnology}\ }\textbf {\bibinfo {volume} {4}},\ \bibinfo {pages}
  {50} (\bibinfo {year} {2009})}\BibitemShut {NoStop}%
\bibitem [{\citenamefont {Shtrikman}\ \emph {et~al.}(2009)\citenamefont
  {Shtrikman}, \citenamefont {Popovitz-Biro}, \citenamefont {Kretinin},
  \citenamefont {Houben}, \citenamefont {Heiblum}, \citenamefont {Buka{\l}a},
  \citenamefont {Galicka}, \citenamefont {Buczko},\ and\ \citenamefont
  {Kacman}}]{Shtrikman2009}%
  \BibitemOpen
  \bibfield  {author} {\bibinfo {author} {\bibfnamefont {H.}~\bibnamefont
  {Shtrikman}}, \bibinfo {author} {\bibfnamefont {R.}~\bibnamefont
  {Popovitz-Biro}}, \bibinfo {author} {\bibfnamefont {A.}~\bibnamefont
  {Kretinin}}, \bibinfo {author} {\bibfnamefont {L.}~\bibnamefont {Houben}},
  \bibinfo {author} {\bibfnamefont {M.}~\bibnamefont {Heiblum}}, \bibinfo
  {author} {\bibfnamefont {M.}~\bibnamefont {Buka{\l}a}}, \bibinfo {author}
  {\bibfnamefont {M.}~\bibnamefont {Galicka}}, \bibinfo {author} {\bibfnamefont
  {R.}~\bibnamefont {Buczko}},\ and\ \bibinfo {author} {\bibfnamefont
  {P.}~\bibnamefont {Kacman}},\ }\bibfield  {title} {\bibinfo {title} {Method
  for suppression of stacking faults in wurtzite iii- v nanowires},\
  }\href@noop {} {\bibfield  {journal} {\bibinfo  {journal} {Nano letters}\
  }\textbf {\bibinfo {volume} {9}},\ \bibinfo {pages} {1506} (\bibinfo {year}
  {2009})}\BibitemShut {NoStop}%
\bibitem [{\citenamefont {Pan}\ \emph {et~al.}(2014)\citenamefont {Pan} \emph
  {et~al.}}]{Pan2014}%
  \BibitemOpen
  \bibfield  {author} {\bibinfo {author} {\bibfnamefont {D.}~\bibnamefont
  {Pan}} \emph {et~al.},\ }\bibfield  {title} {\bibinfo {title} {Controlled
  synthesis of phase-pure inas nanowires on si (111) by diminishing the
  diameter to 10 nm},\ }\href@noop {} {\bibfield  {journal} {\bibinfo
  {journal} {Nano letters}\ }\textbf {\bibinfo {volume} {14}},\ \bibinfo
  {pages} {1214} (\bibinfo {year} {2014})}\BibitemShut {NoStop}%
\bibitem [{\citenamefont {Pientka}\ \emph {et~al.}(2012)\citenamefont
  {Pientka}, \citenamefont {Kells}, \citenamefont {Romito}, \citenamefont
  {Brouwer},\ and\ \citenamefont {Von~Oppen}}]{Brouwer2012ZBP}%
  \BibitemOpen
  \bibfield  {author} {\bibinfo {author} {\bibfnamefont {F.}~\bibnamefont
  {Pientka}}, \bibinfo {author} {\bibfnamefont {G.}~\bibnamefont {Kells}},
  \bibinfo {author} {\bibfnamefont {A.}~\bibnamefont {Romito}}, \bibinfo
  {author} {\bibfnamefont {P.~W.}\ \bibnamefont {Brouwer}},\ and\ \bibinfo
  {author} {\bibfnamefont {F.}~\bibnamefont {Von~Oppen}},\ }\bibfield  {title}
  {\bibinfo {title} {Enhanced zero-bias majorana peak in the differential
  tunneling conductance of disordered multisubband quantum-wire/superconductor
  junctions},\ }\href@noop {} {\bibfield  {journal} {\bibinfo  {journal}
  {Physical review letters}\ }\textbf {\bibinfo {volume} {109}},\ \bibinfo
  {pages} {227006} (\bibinfo {year} {2012})}\BibitemShut {NoStop}%
\bibitem [{\citenamefont {Rainis}\ \emph {et~al.}(2013)\citenamefont {Rainis},
  \citenamefont {Trifunovic}, \citenamefont {Klinovaja},\ and\ \citenamefont
  {Loss}}]{Loss2013ZBP}%
  \BibitemOpen
  \bibfield  {author} {\bibinfo {author} {\bibfnamefont {D.}~\bibnamefont
  {Rainis}}, \bibinfo {author} {\bibfnamefont {L.}~\bibnamefont {Trifunovic}},
  \bibinfo {author} {\bibfnamefont {J.}~\bibnamefont {Klinovaja}},\ and\
  \bibinfo {author} {\bibfnamefont {D.}~\bibnamefont {Loss}},\ }\bibfield
  {title} {\bibinfo {title} {Towards a realistic transport modeling in a
  superconducting nanowire with majorana fermions},\ }\href@noop {} {\bibfield
  {journal} {\bibinfo  {journal} {Physical Review B}\ }\textbf {\bibinfo
  {volume} {87}},\ \bibinfo {pages} {024515} (\bibinfo {year}
  {2013})}\BibitemShut {NoStop}%
\bibitem [{\citenamefont {Pan}\ and\ \citenamefont
  {Das~Sarma}(2020)}]{GoodBadUgly}%
  \BibitemOpen
  \bibfield  {author} {\bibinfo {author} {\bibfnamefont {H.}~\bibnamefont
  {Pan}}\ and\ \bibinfo {author} {\bibfnamefont {S.}~\bibnamefont
  {Das~Sarma}},\ }\bibfield  {title} {\bibinfo {title} {Physical mechanisms for
  zero-bias conductance peaks in majorana nanowires},\ }\href
  {https://doi.org/10.1103/PhysRevResearch.2.013377} {\bibfield  {journal}
  {\bibinfo  {journal} {Phys. Rev. Research}\ }\textbf {\bibinfo {volume}
  {2}},\ \bibinfo {pages} {013377} (\bibinfo {year} {2020})}\BibitemShut
  {NoStop}%
\bibitem [{\citenamefont {Pan}\ \emph {et~al.}(2022)\citenamefont {Pan} \emph
  {et~al.}}]{PanCPL}%
  \BibitemOpen
  \bibfield  {author} {\bibinfo {author} {\bibfnamefont {D.}~\bibnamefont
  {Pan}} \emph {et~al.},\ }\bibfield  {title} {\bibinfo {title} {In situ
  epitaxy of pure phase ultra-thin inas-al nanowires for quantum devices},\
  }\href {https://doi.org/10.1088/0256-307X/39/5/058101} {\bibfield  {journal}
  {\bibinfo  {journal} {Chinese Physics Letters}\ }\textbf {\bibinfo {volume}
  {39}},\ \bibinfo {eid} {058101} (\bibinfo {year} {2022})}\BibitemShut
  {NoStop}%
\bibitem [{\citenamefont {Wang}\ \emph {et~al.}(2023)\citenamefont {Wang} \emph
  {et~al.}}]{Zhichuan}%
  \BibitemOpen
  \bibfield  {author} {\bibinfo {author} {\bibfnamefont {Z.}~\bibnamefont
  {Wang}} \emph {et~al.},\ }\bibfield  {title} {\bibinfo {title} {Supercurrent
  in a quasi-ballistic thin inas-al hybrid nanowire device},\ }\href@noop {}
  {\bibfield  {journal} {\bibinfo  {journal} {to be appeared}\ } (\bibinfo
  {year} {2023})}\BibitemShut {NoStop}%
\bibitem [{\citenamefont {Reed}\ \emph {et~al.}(2010)\citenamefont {Reed},
  \citenamefont {DiCarlo}, \citenamefont {Johnson}, \citenamefont {Sun},
  \citenamefont {Schuster}, \citenamefont {Frunzio},\ and\ \citenamefont
  {Schoelkopf}}]{PRL_Reed}%
  \BibitemOpen
  \bibfield  {author} {\bibinfo {author} {\bibfnamefont {M.~D.}\ \bibnamefont
  {Reed}}, \bibinfo {author} {\bibfnamefont {L.}~\bibnamefont {DiCarlo}},
  \bibinfo {author} {\bibfnamefont {B.~R.}\ \bibnamefont {Johnson}}, \bibinfo
  {author} {\bibfnamefont {L.}~\bibnamefont {Sun}}, \bibinfo {author}
  {\bibfnamefont {D.~I.}\ \bibnamefont {Schuster}}, \bibinfo {author}
  {\bibfnamefont {L.}~\bibnamefont {Frunzio}},\ and\ \bibinfo {author}
  {\bibfnamefont {R.~J.}\ \bibnamefont {Schoelkopf}},\ }\bibfield  {title}
  {\bibinfo {title} {High-fidelity readout in circuit quantum electrodynamics
  using the jaynes-cummings nonlinearity},\ }\href
  {https://doi.org/10.1103/PhysRevLett.105.173601} {\bibfield  {journal}
  {\bibinfo  {journal} {Phys. Rev. Lett.}\ }\textbf {\bibinfo {volume} {105}},\
  \bibinfo {pages} {173601} (\bibinfo {year} {2010})}\BibitemShut {NoStop}%
\bibitem [{\citenamefont {Purcell}\ \emph {et~al.}(1946)\citenamefont
  {Purcell}, \citenamefont {Torrey},\ and\ \citenamefont {Pound}}]{Purcell}%
  \BibitemOpen
  \bibfield  {author} {\bibinfo {author} {\bibfnamefont {E.~M.}\ \bibnamefont
  {Purcell}}, \bibinfo {author} {\bibfnamefont {H.~C.}\ \bibnamefont
  {Torrey}},\ and\ \bibinfo {author} {\bibfnamefont {R.~V.}\ \bibnamefont
  {Pound}},\ }\bibfield  {title} {\bibinfo {title} {Resonance absorption by
  nuclear magnetic moments in a solid},\ }\href
  {https://doi.org/10.1103/PhysRev.69.37} {\bibfield  {journal} {\bibinfo
  {journal} {Phys. Rev.}\ }\textbf {\bibinfo {volume} {69}},\ \bibinfo {pages}
  {37} (\bibinfo {year} {1946})}\BibitemShut {NoStop}%
\bibitem [{\citenamefont {Schuster}\ \emph {et~al.}(2007)\citenamefont
  {Schuster}, \citenamefont {Houck}, \citenamefont {Schreier}, \citenamefont
  {Wallraff}, \citenamefont {Gambetta}, \citenamefont {Blais}, \citenamefont
  {Frunzio}, \citenamefont {Majer}, \citenamefont {Johnson}, \citenamefont
  {Devoret}, \citenamefont {Girvin},\ and\ \citenamefont
  {Schoelkopf}}]{Photon_number}%
  \BibitemOpen
  \bibfield  {author} {\bibinfo {author} {\bibfnamefont {D.}~\bibnamefont
  {Schuster}}, \bibinfo {author} {\bibfnamefont {A.}~\bibnamefont {Houck}},
  \bibinfo {author} {\bibfnamefont {J.}~\bibnamefont {Schreier}}, \bibinfo
  {author} {\bibfnamefont {A.}~\bibnamefont {Wallraff}}, \bibinfo {author}
  {\bibfnamefont {J.}~\bibnamefont {Gambetta}}, \bibinfo {author}
  {\bibfnamefont {A.}~\bibnamefont {Blais}}, \bibinfo {author} {\bibfnamefont
  {L.}~\bibnamefont {Frunzio}}, \bibinfo {author} {\bibfnamefont
  {J.}~\bibnamefont {Majer}}, \bibinfo {author} {\bibfnamefont
  {B.}~\bibnamefont {Johnson}}, \bibinfo {author} {\bibfnamefont
  {M.}~\bibnamefont {Devoret}}, \bibinfo {author} {\bibfnamefont
  {S.}~\bibnamefont {Girvin}},\ and\ \bibinfo {author} {\bibfnamefont
  {R.}~\bibnamefont {Schoelkopf}},\ }\bibfield  {title} {\bibinfo {title}
  {Resolving photon number states in a superconducting circuit},\ }\href
  {https://doi.org/10.1038/nature05461} {\bibfield  {journal} {\bibinfo
  {journal} {Nature}\ }\textbf {\bibinfo {volume} {445}},\ \bibinfo {pages}
  {515} (\bibinfo {year} {2007})}\BibitemShut {NoStop}%
\bibitem [{\citenamefont {Krantz}\ \emph {et~al.}(2019)\citenamefont {Krantz},
  \citenamefont {Kjaergaard}, \citenamefont {Yan}, \citenamefont {Orlando},
  \citenamefont {Gustavsson},\ and\ \citenamefont {Oliver}}]{guide}%
  \BibitemOpen
  \bibfield  {author} {\bibinfo {author} {\bibfnamefont {P.}~\bibnamefont
  {Krantz}}, \bibinfo {author} {\bibfnamefont {M.}~\bibnamefont {Kjaergaard}},
  \bibinfo {author} {\bibfnamefont {F.}~\bibnamefont {Yan}}, \bibinfo {author}
  {\bibfnamefont {T.~P.}\ \bibnamefont {Orlando}}, \bibinfo {author}
  {\bibfnamefont {S.}~\bibnamefont {Gustavsson}},\ and\ \bibinfo {author}
  {\bibfnamefont {W.~D.}\ \bibnamefont {Oliver}},\ }\bibfield  {title}
  {\bibinfo {title} {A quantum engineer's guide to superconducting qubits},\
  }\href {https://doi.org/10.1063/1.5089550} {\bibfield  {journal} {\bibinfo
  {journal} {Applied Physics Reviews}\ }\textbf {\bibinfo {volume} {6}},\
  \bibinfo {pages} {021318} (\bibinfo {year} {2019})}\BibitemShut {NoStop}%
\end{thebibliography}%

\newpage

\onecolumngrid

\newpage


\section*{Supplementary material}

# Supplementary Information for "Gatemon qubit based on a thin InAs-Al hybrid nanowire"

Jierong Huo,[1,*] Zezhou Xia,[1,*] Zonglin Li,[1,*] Shan Zhang,[1,*] Yuqing Wang,[2] Dong Pan,[3] Qichun Liu,[2] Yulong Liu,[2] Zhichuan Wang,[4] Yichun Gao,[1] Jianhua Zhao,[3] Tiefu Li,[5,2] Jianghua Ying,[6,†] Runan Shang,[2] and Hao Zhang[1,2,7,‡]

[1]*State Key Laboratory of Low Dimensional Quantum Physics, Department of Physics, Tsinghua University, Beijing 100084, China*
[2]*Beijing Academy of Quantum Information Sciences, Beijing 100193, China*
[3]*State Key Laboratory of Superlattices and Microstructures, Institute of Semiconductors, Chinese Academy of Sciences, P. O. Box 912, Beijing 100083, China*
[4]*Beijing National Laboratory for Condensed Matter Physics, Institute of Physics, Chinese Academy of Sciences, Beijing 100190, China*
[5]*School of Integrated Circuits and Frontier Science Center for Quantum Information, Tsinghua University, Beijing 100084, China*
[6]*Yangtze Delta Region Industrial Innovation Center of Quantum and Information, Suzhou 215133, China*
[7]*Frontier Science Center for Quantum Information, Beijing 100084, China*

## Method

### Resonator fabrication

A sapphire substrate was covered by a 100-nm-thick NbTiN superconducting film using sputtering. The NbTi target (Nb/Ti 70/30 Wt%, 99.95% Pure) was sputtered at a pressure of 3.5 mTorr with a gas mixture Ar:$N_2$ = 18:1. Then direct laser writing was used to define the patterns for the feed line, the cavity, the T-shape capacitor and the substrate region for InAs-Al contacts and gates. The photo-resist was S1813 (3000 rpm, 60 s, 115 °C for 120 s). After developing in AZ for 1 min, reaction ion etching ($O_2$ 5 Pa for 20 s, $CF_4$ 2 Pa for 135 s) was performed to etch away the exposed regions of the NbTiN film. Finally, the residual resist was removed in acetone.

### Qubit fabrication

The thin InAs nanowires were grown by molecular-beam epitaxy followed by an in-situ growth of the Al film (half shell). The hybrid wires were then transferred from the growth chip onto the resonator chip by wipes of clean room tissues. PMMA 672.045 resist and photoresist AR-PC 5090.02 were spun at 4000 rpm for 1 min and baked at 120 °C for 10 min and 90 °C for 2 min, respectively. Electron beam lithography (EBL) was performed to pattern the etch windows. After development in MIBK:IPA = 1:3 for 50 s, the chip was immersed in Transene Aluminum Etchant Type D at 50 °C for 10 s to etch way the exposed Al shells. Another EBL was performed for the contacts and the side gate electrodes by sputtering Ti/NbTiN (1/100 nm). Before the sputtering, a short argon plasma etching (90 s, 50 W, 0.05 Torr) was performed to ensure Ohmic contact.

---

[*] equal contribution
[†] yingjianghua@tgqs.net
[‡] hzquantum@mail.tsinghua.edu.cn

2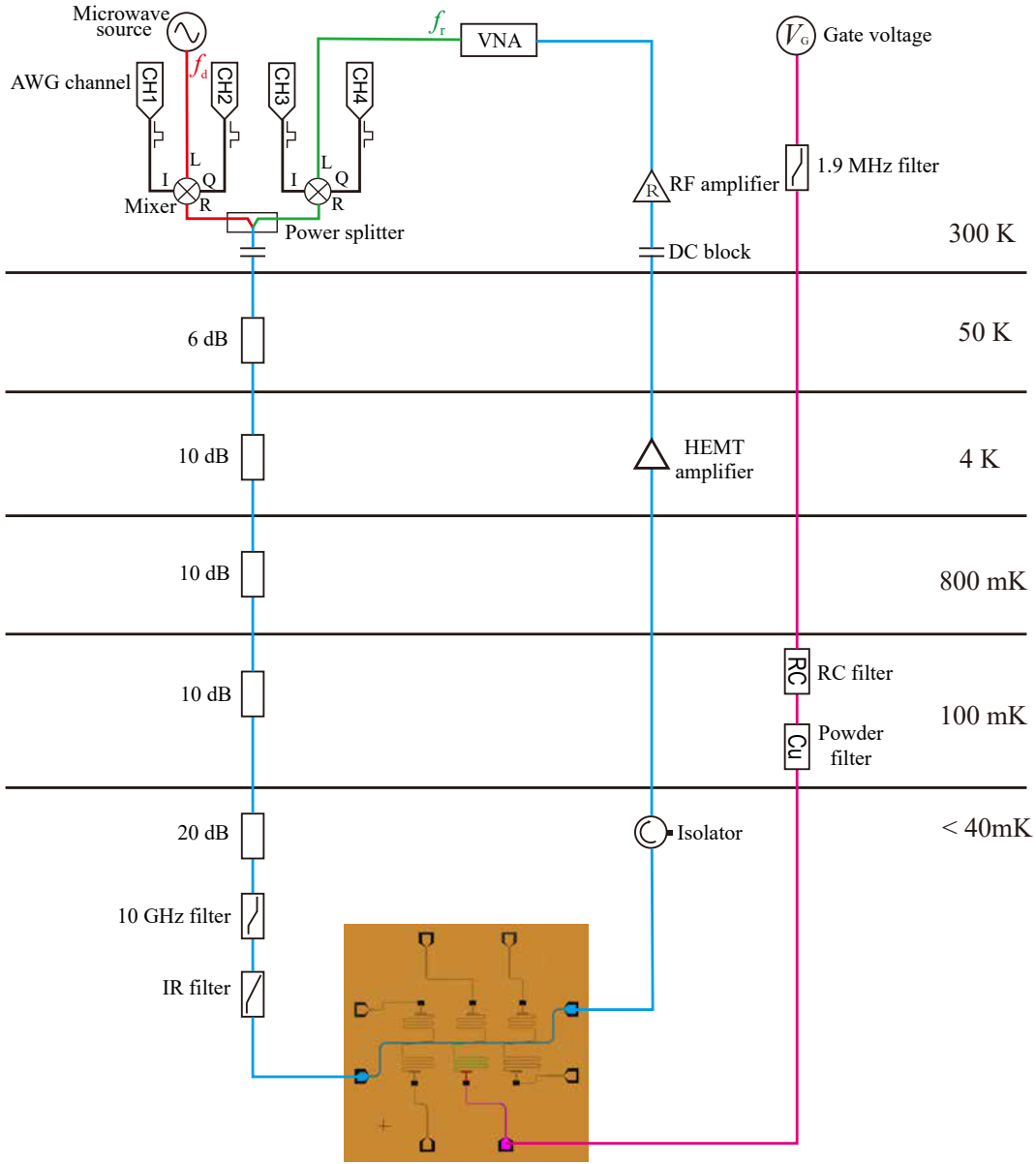

FIG. S1. Gatemon measurement circuit. The red line is for the qubit control. A continuous signal (frequency $f_\mathrm{d}$) generated from the microwave source was added into the mixer (the L channel). The I and Q channels of the mixer were square wave pulses from an arbitrary waveform generator (AWG). After the mixer, a drive pulse (frequency $f_\mathrm{d}$) of designed duration time can be generated to excite and control the qubit. The green line is for the qubit readout. A microwave signal (frequency $f_\mathrm{r}$) generated from the VNA was added into another mixer. Similarly, square wave pulses from the AWG modulates the time of the readout pulse through the mixer (for the continuous method, no mixer is needed). $f_\mathrm{r}$ was tuned close to the cavity resonance frequency $f_\mathrm{C}$. These two microwave tones were added together and then were fed into one end of the feed line on the chip after passing several cryogenic attenuators and filters (for noise suppression). The transmission ($S_{21}$) of the readout pulse was measured at the other end of the feed line by the VNA. The isolator and the high electron mobility transistors (HEMT) amplifier protects the sample from thermal noise and photons. The pink line is for the gate line to apply a DC voltage.

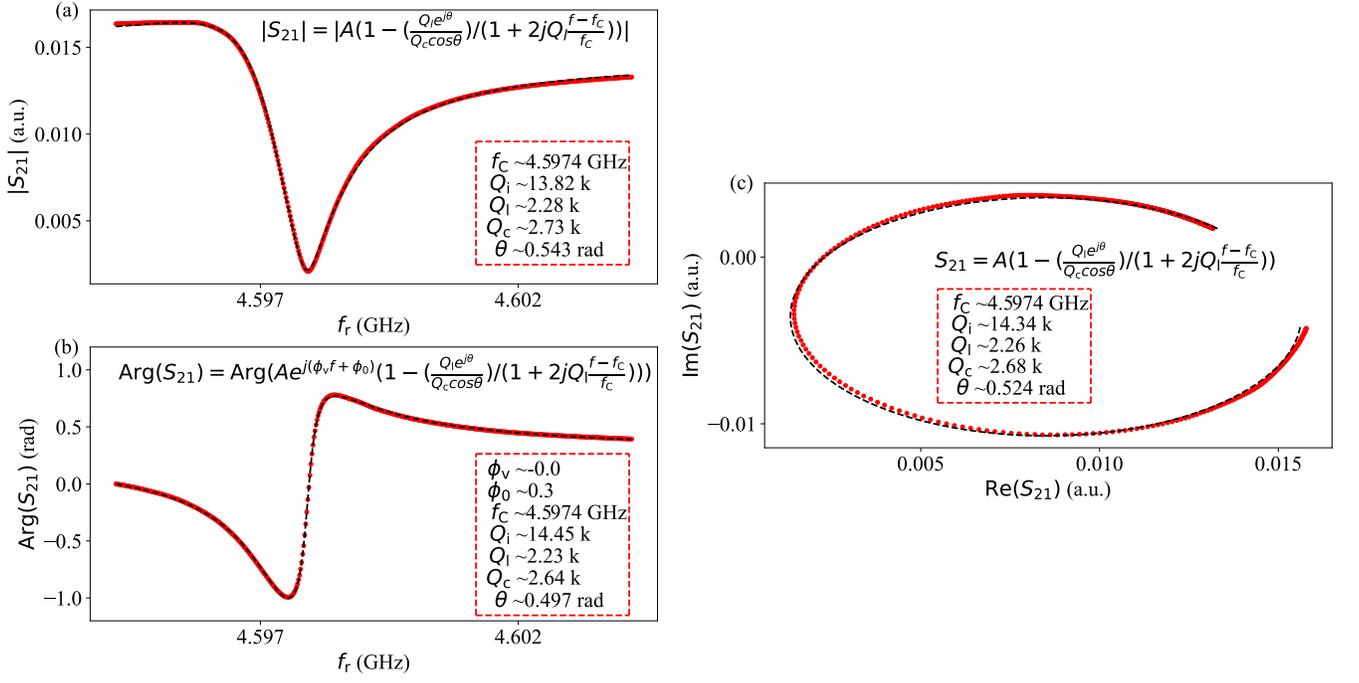

FIG. S2. Extraction of the cavity quality factor. (a) The cavity transmission amplitude $|S_{21}|$ as a function of the readout frequency. The fitting curve (black dashed line) uses the function: $S_{21} = Ae^{j(\phi_\nu f + \phi_0)}[1 - (\frac{Q_l e^{j\theta}}{Q_c \cos(\theta)})/(1 + 2jQ_l \frac{f-f_C}{f_C})]$. The terms $A$, $\phi_\nu$, $\phi_0$, $Q_l$, $Q_c$ and $\theta$ are the amplitude of the transmission off-resonance, the constant phase propagation due to the length of cables, the offset in the phase, the load quality factor, the coupling quality factor and the parameter to describe asymmetry of the Lorentzian line shape, respectively. We extract the internal quality factor $Q_i = (Q_l^{-1} - Q_c^{-1})^{-1} \sim 13.8$k. (b) Fitting the phase, $\text{Arg}(S_{21})$. We estimate $Q_i \sim 14.45$k. (c) $\text{Im}\{S_{21}\}$ vs. $\text{Re}\{S_{21}\}$ and the least-squres fit. The term $e^{j(\phi_\nu f + \phi_0)}$ has been incorporated into $S_{21}$.

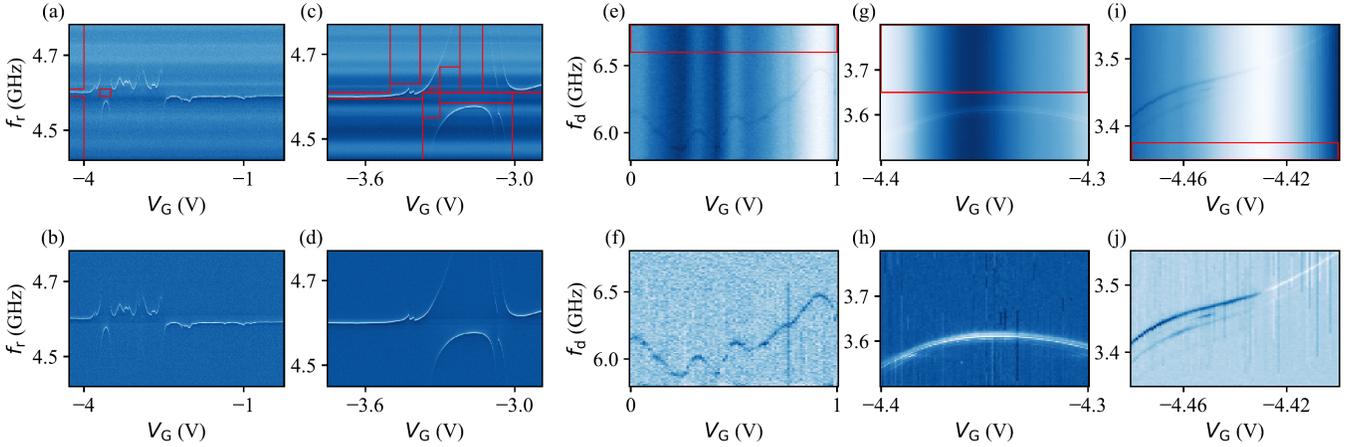

FIG. S3. Background subtraction in Fig. 2. (a) Raw data of Fig. 2(a). We average the signal (along $V_G$) in the red boxes as the background signal, likely due to standing waves in the circuit. No cavity signal is involved in this averaging method. (b) The spectroscopy after the background subtraction. (c), (e), (g) and (i) are the raw data of Figs. 2(b) and 2(e), while (d), (f), (h) and (j) are the versions after the background subtraction. For the two-tone spectroscopy, the background signal is due to different working points selected at different $V_G$'s.

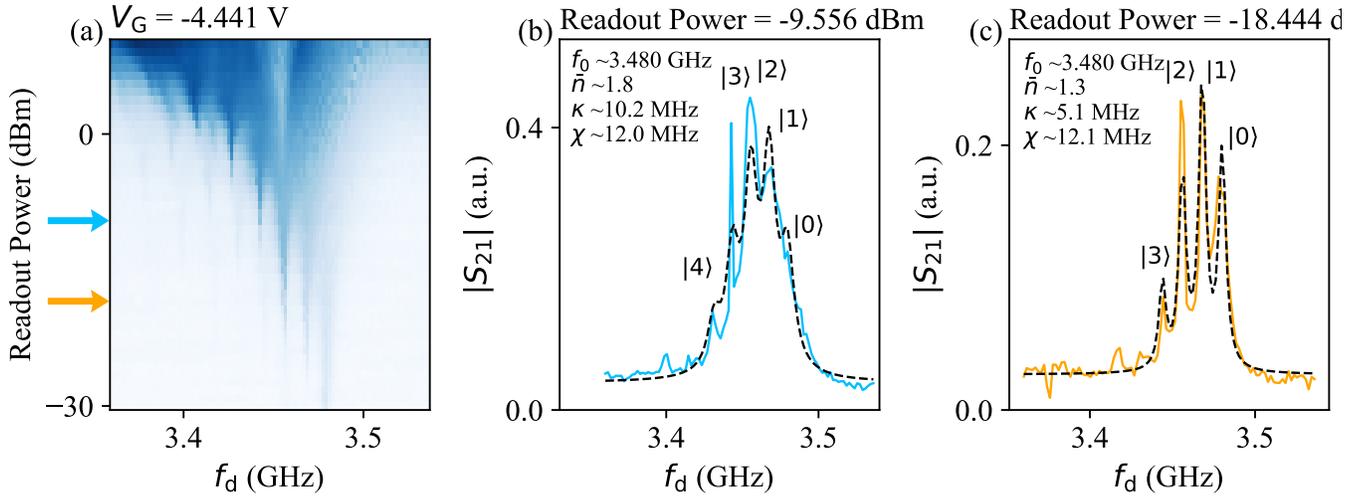

FIG. S4. (a) Photon-number-dependent frequency shift as a function of the readout power. (b) and (c) are the line cuts of (a). The fitting (dashed lines) is based on Lorentzian peaks with the Poisson distribution. The deviations are likely due to the thermal effect.

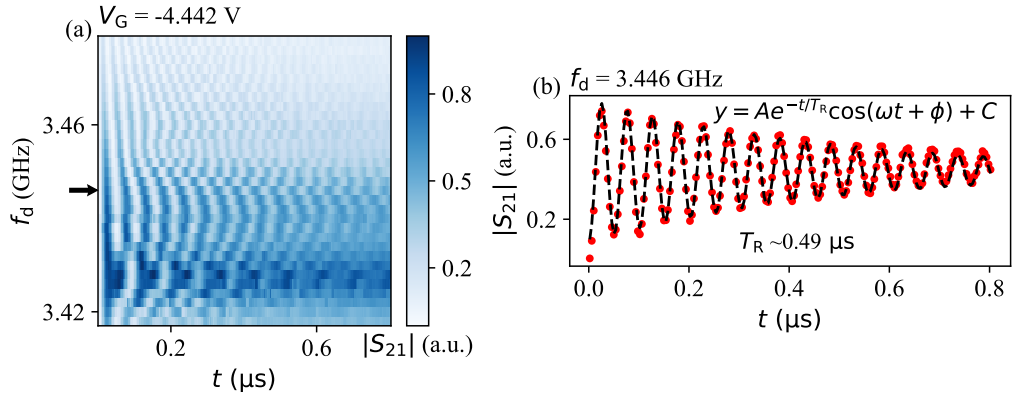

FIG. S5. (a) Rabi oscillations at $V_G = -4.442$ V. (b) A line cut (see the arrow in (a)) with the exponential fit (dashed line).

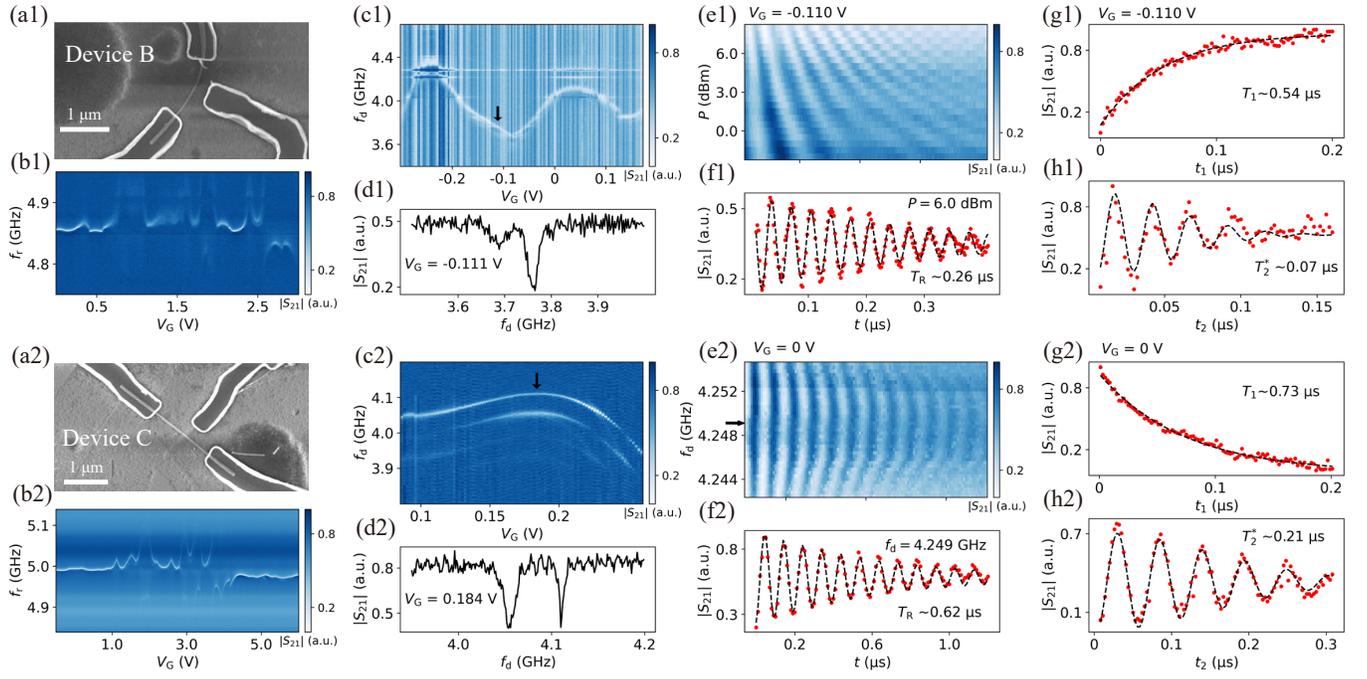

FIG. S6. Two more gatemon qubits. (a1)-(h1) for device B. (a1) SEM of the Josephson junction region. (b1) Cavity spectroscopy (single tone). (c1) Two-tone qubit spectroscopy. (d1) A line cut from (c1). (e1) Rabi oscillations. (f1) A line cut from (e1). (g1) $T_1$ measurement. (h1) $T_2^*$ measurement. (a2)-(h2) similar to (a1)-(h1) but for device C.



\end{document}